\documentclass[fleqn,usenatbib]{mnras}

\usepackage[T1]{fontenc}
\usepackage{ae,aecompl}

\usepackage{graphicx}	
\usepackage{amsmath}	
\usepackage{amssymb}	

\usepackage{color}

\newcommand{\ron}{\color{red}}

\title{WD 1202-024: The Shortest-Period Pre-Cataclysmic Variable}

\author[Rappaport et al.]{
S.~Rappaport$^1$,   
A.~Vanderburg$^2$, 
L.~Nelson$^3$, 
B.L.~Gary$^4$, 
T.G.~Kaye$^5$, 
B.~Kalomeni$^6$, 
\newauthor
S.B.~Howell$^7$, 
J.R.~Thorstensen$^8$, 
F.-R. Lachapelle$^{9}$,
M.~Lundy$^3$,
J.~St-Antoine$^{9}$
\\
$^{1}$ Department of Physics, and Kavli Institute for Astrophysics and Space Research, M.I.T., Cambridge, MA 02139, USA; sar@mit.edu \\
$^{2}$ Harvard-Smithsonian Center for Astrophysics, 60 Garden Street, Cambridge, MA 02138 USA; avanderburg@cfa.harvard.edu \\
$^{3}$ Department of Physics and Astronomy, Bishop's University, 2600 College St., Sherbrooke, QC J1M 1Z7; lnelson@ubishops.ca \\ 
$^{4}$ Hereford Arizona Observatory, Hereford, AZ 85615; BLGary@umich.edu \\
$^{5}$ Raemor Vista Observatory, 7023 E. Alhambra Dr., Sierra Vista, AZ 85650; tom@TomKaye.com \\
$^{6}$ Department of Astronomy and Space Sciences, Ege University, 35100, \.Izmir, Turkey; belinda.kalomeni@ege.edu.tr \\
$^{7}$ Space Science \& Astrobiology Division, NASA Ames Research Center, M/S 245-1, Moffett Field, CA 94035; steve.b.howell@nasa.gov \\ 
$^{8}$ Department of Physics and Astronomy, Dartmouth College, 239 Wilder Hall, Hanover, NH 03755; john.r.thorstensen@dartmouth.edu \\
$^{9}$ Institut de recherche sur les exoplan\'etes (iREx),D\'epartement de Physique, Universit\'e de Montr\'eal, Montr\'eal, QC  H3C 3J7 \\
}

\date{Submitted 2017 February 28}

\pubyear{2017}

\begin{document}
\label{firstpage}
\pagerange{\pageref{firstpage}--\pageref{lastpage}}
\maketitle

\begin{abstract}
Among the 28,000 targeted stars in K2 Field 10 is the white dwarf WD 1202-024 (EPIC 201283111), first noted in the SDSS survey (SDSS 120515.80-024222.7).  We have found that this hot white dwarf ($T_{\rm eff} \simeq 22,640$ K) is in a very close orbit ($P \simeq 71$ min) with a star of near brown-dwarf mass $\simeq 0.061 \, M_\odot$.  This period is very close to, or somewhat below, the minimum orbital period of cataclysmic variables with H-rich donor stars.  However, we find no evidence that this binary is currently, or ever was, transferring mass from the low-mass companion to the white dwarf.  We therefore tentatively conclude that this system is still in the pre-cataclysmic variable phase, having emerged from a common envelope some $50 \pm 20$ Myr ago.  Because of the 29-minute integration time of K2, we use follow-up ground-based photometry to better evaluate the eclipsing light curve.  We also utilize the original SDSS spectra, in approximately 15-min segments, to estimate the radial velocity of the white dwarf in its orbit.  An analysis of the light curve, with supplementary constraints, leads to the following system parameters: $M_{\rm wd} \simeq 0.415 \pm 0.028 \, M_\odot$, $R_{\rm wd} \simeq 0.021 \pm 0.001 \, R_\odot$, $M_{\rm com} \simeq 0.061 \pm 0.010 \, M_\odot$, and $R_{\rm com} \simeq 0.088 \pm 0.005 \, R_\odot$ where the subscripts `wd' and `com' refer to the white dwarf and low-mass companion respectively. If our interpretation of this system as a pre-CV is correct, it has the shortest period of any such system yet found and should become a compact CV in less than 250 Myr.
\end{abstract}

\begin{keywords}
stars: binaries: eclipsing -- binaries: general -- stars: dwarf novae -- novae, cataclysmic variables -- white dwarfs
\end{keywords}



\section{Introduction}
\label{sec:intro}

The standard scenario for the evolution of cataclysmic variable (`CV') binaries is discussed in numerous papers, including Rappaport et al.~(1983), Patterson (1984), Warner (1995), Howell et al.~(2001), Knigge et al.~(2011), Kalomeni et al.~(2016), and references therein.  The formation of a CV starts when the more massive star in the primordial binary evolves first, and, in the process, develops a high-density, hydrogen exhausted core.  When the primary expands up the giant or asymptotic giant branch, depending on the initial orbital separation, it may eventually overflow its Roche lobe and commence mass transfer onto the companion star.  Depending on the mass ratio of the binary, and the evolutionary state of the primary, the mass transfer may be dynamically unstable, leading to a so-called `common-envelope' phase (see, e.g., Paczy\'nski 1976; Taam et al. 1978; Webbink 1984; Taam \& Bodenheimer 1992; Pfahl et al.~2003).  In this scenario, the secondary spirals into the envelope of the primary.  If there is sufficient potential energy between the core of the primary and the secondary star, the entire envelope of the primary can be ejected, leaving a hot proto-white dwarf (i.e., the core of the primary) in a fairly short period orbit with the original secondary.

The result of this scenario is a phase that is often referred to as a `post-common-envelope binary' or a `pre-cataclysmic variable' (`pre-CV').  At the present time there are more than 60 such objects known (see, e.g., Zorotovic et al.~2011), many of which were initially identified via the SDSS survey.  The companion stars have masses that are mostly in the range of $0.1-0.7 \,M_\odot$ and orbital periods of $\sim$2 hours to a day.  The white dwarfs consist of a mix of He WDs (with masses of $\sim$$0.3-0.5 \, M_\odot$) and CO WDs (masses $\gtrsim 0.5 M_\odot$).

If and when the companion star itself evolves to fill its Roche lobe, and/or the orbit shrinks due to angular momentum losses (e.g., gravitational radiation or magnetic braking; Rappaport et al.~1983), mass transfer onto the white dwarf can commence, thereby marking the zero-age cataclysmic variable (`ZACV') phase.  If the mass transfer during the initial phases of the CV is stable, then there are three known possible outcomes of the evolution depending on the initial conditions: (i) evolution to wide orbits where the donor star becomes a giant; (ii) evolution to shorter periods, where a minimum period of $\sim$72-75 minutes is reached, after which the orbit expands; and (iii) evolution to ultrashort periods of $\sim$5 minutes after which the orbital period increases (see, e.g., Kalomeni et al.~2016 for an extensive review).  

The minimum orbital period in a H-rich CV occurs when the donor star becomes degenerate at a mass of $\sim$0.05 $M_\odot$, and starts to expand upon further mass loss.  Thus, at a period of $\sim$70 minutes a donor star that may have had an initial mass as high as 2 $M_\odot$ can have been reduced to a mass of brown-dwarf proportions with most of its initial mass having been lost from the system.  Therefore, when we see a system with a period of $\sim$70 minutes with a brown-dwarf like companion, the important issue arises as to whether the low-mass companion star has been whittled down from a much larger star, or if the system has only recently emerged from a common envelope with the star that ejected the envelope being relatively unscathed.  In the latter case, it would be interesting to understand whether the common envelope phase can end with the companion star nearly filling its Roche lobe without actually doing so.

In this paper we report on a binary system, WD 1202-024, which may indeed have only recently emerged from a common-envelope phase, and is not yet in Roche-lobe contact.  In Sect.~\ref{sec:K2} we describe the K2 discovery and observations of WD 1202-024.  In Sect.~\ref{sec:SDSS}, we utilize a published analysis of an SDSS spectrum of WD 1202-024 obtained in 2002 to establish both $T_{\rm eff}$ and $\log \, g$ for the white dwarf.  We also carry out a radial velocity analysis by utilizing the three individual exposures contributing to the SDSS spectrum.  Our ground-based photometry from five observatories is reported in Sect.~\ref{sec:photometry}.  The system parameters are inferred in Sect.~\ref{sec:model} from an analysis of the lightcurve coupled with other constraints on the system.  In Sect.~\ref{sec:origin} we discuss the evolutionary status of WD 1202-024, and possible evolutionary scenarios to explain its origin.  We summarize and discuss our results in Sect.~\ref{sec:concl}.

\section{K2 Observations}
\label{sec:K2}

\begin{figure}
\begin{center}
\includegraphics[width=1.01 \columnwidth]{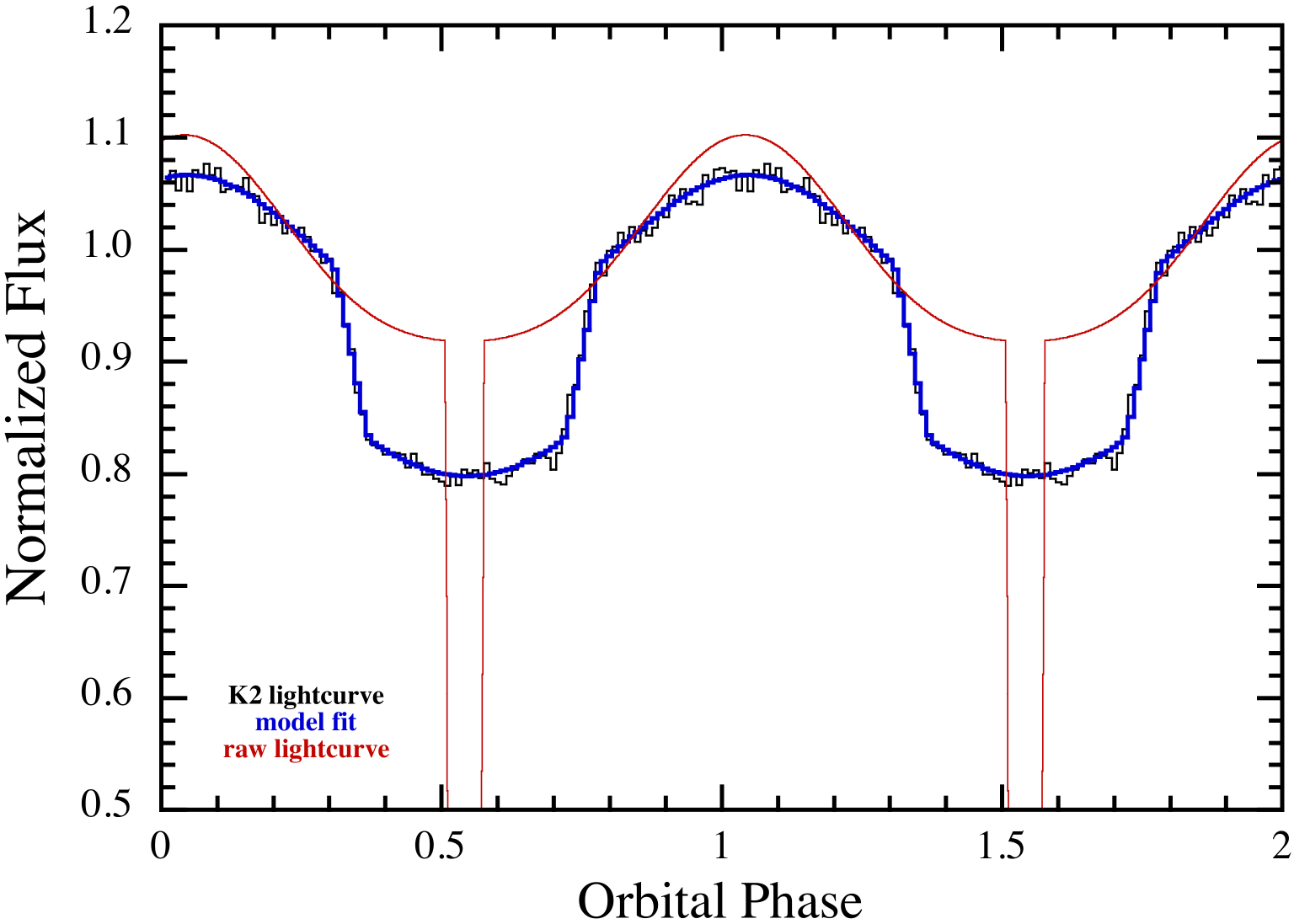}
\caption{K2 lightcurve of WD 1202-024 folded about a period 71.2299-minutes. (Phase zero is offset for aesthetic reasons.)  The red curve represents a simple geometrical model with a 5-minute long eclipse and two out-of-eclipse cosine functions to emulate the illumination effect on the secondary star--with a $\sim$18\% peak-to-peak amplitude.  The heavy blue curve is that model convolved with the K2 integration time of 29.4 minutes.}
\label{fig:lightcurve} 
\end{center}
\end{figure}

In our continuing search for unusual eclipsing binaries in the K2 observations (Howell et al.~2014), we downloaded all available K2 Extracted Lightcurves common to Campaign 10 from the MAST\footnote{\url{http://archive.stsci.edu/k2/data\_search/search.php}}. We utilized the pipelined data set of Vanderburg \& Johnson (2014).  The flux data from all 28,345 targets were searched for periodicities via a BLS algorithm (Kov\'acs et al.~2002) as well as via an FFT search.  Folded lightcurves of targets with significant peaks in their periodograms were then examined by eye to identify unusual objects among those with distinct periodicities.  

Shortly after the release of the Field 10 data set, EPIC 201283111 (WD 1202-024) was identified as a very short 71-minute period binary with a single discernible eclipse per orbit.  The folded K2 lightcurve for WD 1202-024 is shown in Fig.~\ref{fig:lightcurve}.   

Because the K2 integration time is a significant fraction (41\%) of the orbital period, the lightcurve shown in Fig.~\ref{fig:lightcurve} is highly smoothed.  In fact, the eclipse width appears to be roughly 30 minutes in duration, indicating that the actual eclipse is much shorter.  The figure also shows a simple geometrical `toy' model (red curve) with a 5-minute long rectangular shaped eclipse, as well as out-of-eclipse cosine functions (at one and two times the orbital frequency; see Sect.~\ref{sec:MCMC}) to emulate the illumination effect on the secondary star--with an $\sim$18\% peak-to-peak amplitude. The heavy blue curve is that same model convolved with the K2 integration time of 29.4 minutes, and it matches the observed light curve quite well.

\begin{table}
\centering
\caption{Photometric and Spectral Properties of WD 1202-024}
\begin{tabular}{lc}
\hline
\hline
Parameter &
WD 1202-024 \\
\hline
RA (J2000) & 12:05:15.808   \\  
Dec (J2000) &  $-02$:42:22.69  \\  
$K_p$ & 18.75  \\
$u^a$ &  18.50   \\
$B^b$ & 18.51 \\  
$g^a$ & 18.49  \\
$V^b$ & 18.67 \\  
$r^a$ & 18.84  \\  
$i^a$ & 19.15 \\
$z^a$ & 19.39  \\
$T_{\rm eff}$$^c$ (K) & $22,640 \pm 540$ \\
$\log \, g$$^c$ (cgs) & $7.30 \pm 0.10$ \\
Distance (pc)$^d$ & $845 \pm 60$  \\   
$\mu_\alpha$ (mas ~${\rm yr}^{-1}$)$^e$ & $-18.0 \pm 1.4$  \\ 
$\mu_\delta$ (mas ~${\rm yr}^{-1}$)$^e$ &  $+8.0 \pm 1.7$  \\ 
\hline
\label{tbl:mags}
\end{tabular}

{\bf Notes.} (a) Taken from the SDSS image (Ahn et al.~2012). (b) Converted from the SDSS magnitudes; see also VizieR \url{http://vizier.u-strasbg.fr/}; UCAC4 (Zacharias et al.~2013).  (c) From Kleinman et al.~(2013).  (d)  Based on photometric parallax only.  (e) From Qi et al.~(2015). 
\end{table}

\section{SDSS Data Archive}
\label{sec:SDSS}

\begin{figure}
\begin{center}
\includegraphics[width=0.99 \columnwidth]{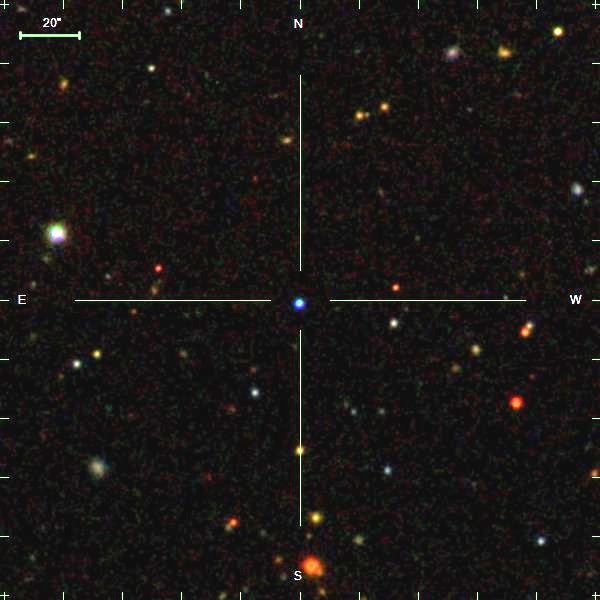} 
\caption{SDSS image of WD 1202-024 showing the 19th magnitude white dwarf well isolated from neighboring stars of comparable brightness.}
\label{fig:SDSSimage} 
\end{center}
\end{figure}

Fortunately, WD 1202-024 was observed by the Sloan Digital Sky Survey (Eisenstein et al.~2006; Ahn et al.~2012) 15 years ago in April of 2002.  An SDSS image is presented in Fig.~\ref{fig:SDSSimage} and shows a 19th magnitude blue star identified as a hot WD via the SDSS spectrum.  There are no obvious nearby companions of comparable brightness.  

WD 1202-024 was first reported as a white dwarf by Eisenstein et al.~(2006) with $T_{\rm eff} \simeq 22,280 \pm 530$ K and $\log \, {\rm g} \simeq 7.18 \pm 0.09$.  This was followed by Kleinman et al.~(2013) who reported $T_{\rm eff} \simeq 22,640 \pm 540$ K and $\log \, {\rm g} \simeq 7.30 \pm 0.10$ as part of their survey of $\sim$20,000 white dwarfs in SDSS DR7.  We adopt this more recently analyzed set of parameters for our analysis.  The photometric and spectral properties of WD 1202-024 are summarized in Table \ref{tbl:mags}, and we note that the light from this system is totally dominated by the white dwarf.

\begin{figure*}
\begin{center}
\includegraphics[width=0.65 \textwidth]{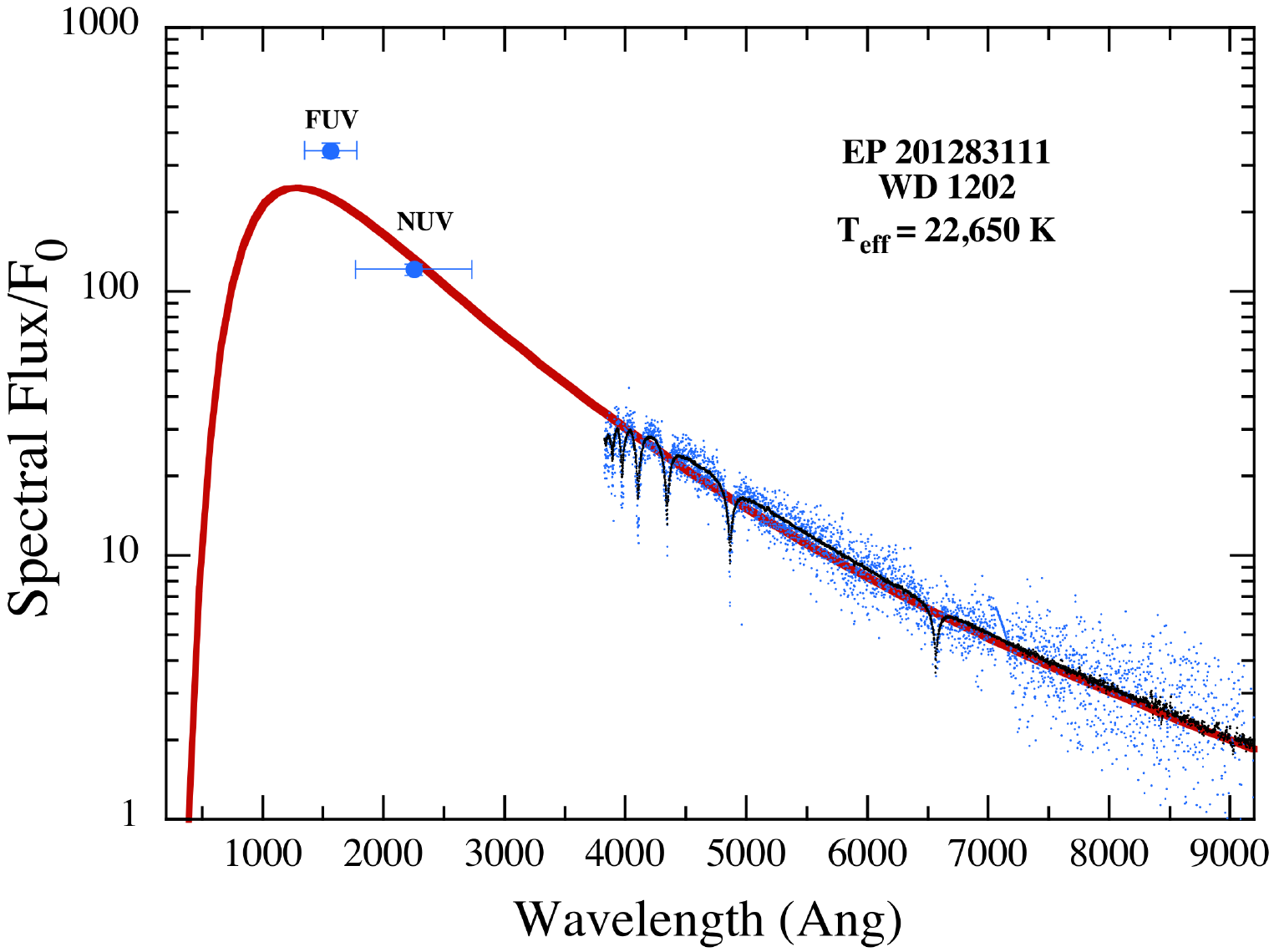}  
\caption{SDSS spectrum of WD 1202-024 (blue points) showing prominent Balmer lines. The normalization constant $F_0$ is $10^{-17}$ ergs cm$^{-2}$ s$^{-1}$ \AA $^{-1}$.  The {\em Galex} FUV and NUV spectral points are also shown in blue.  The black curve is a model spectrum over the visible band.  The red curve is a blackbody of $T_{\rm eff} = 22,650$ K whose amplitude has been fitted to the data, and is meant only to guide the eye.} 
\label{fig:SDSSspectrum} 
\end{center}
\end{figure*}

The SDSS spectrum of this object has a total exposure of about 50 minutes, and is presented in Fig.~\ref{fig:SDSSspectrum}.  Also shown on the same spectral plot are the FUV and NUV flux points from the {\em Galex} satellite\footnote{\url{http://galex.stsci.edu/GR6/}} (Martin et al.~2005).  We also superpose a blackbody curve for $T_{\rm eff} = 22,650$ K to guide the eye.  According to Vennes et al.~(2011) the observed color magnitudes of this white dwarf, FUV-NUV $= -0.32 \pm 0.08$ and NUV-V $=-0.51 \pm 0.07$, imply a white dwarf temperature of $\sim$19,600 K.  When the colors are adjusted for interstellar extinction (according to the prescriptions of Cardelli et al.~1989), assuming a minimal value for E(B-V) of $\simeq 0.03$, they yield a corrected $T_{\rm eff}$ for the white dwarf of $\simeq 22,000$ K, thereby roughly confirming the value found spectroscopically by Kleinman et al.~(2013).

We downloaded the SDSS spectral data which consisted of 3 integrations of 15, 15, and 20 minute durations.  The times and exposures of the spectra are summarized in Table \ref{tbl:RVs}.  We performed two analyses to measure the radial velocity of the three sub-exposures of the SDSS spectrum.  We first performed a cross-correlation function (`CCF') analysis to measure radial velocities.  After excluding outliers, we cross-correlated the three individual spectra against the model spectrum selected by the SDSS pipeline and recorded the velocity of the cross-correlation peak as the radial velocity.  In addition to the CCF analysis, we measured radial velocities by modeling the stellar spectrum with a radial velocity shift (and two nuisance parameters for continuum offset and sky background contamination), and finding the most likely radial velocity shift.  We used a Markov-chain Monte Carlo (`MCMC') algorithm (with affine invariant ensemble sampling; Goodman \& Weare 2010; Hou et al.~2012) to explore parameter space and determine the best-fit velocity shift and uncertainties. The resultant radial velocities ('RVs') that were determined in this manner are summarized in Table \ref{tbl:RVs}.   

The SDSS spectra are of low signal-to-noise, and the white dwarf spectrum has only broad hydrogen absorption lines, so the radial velocity uncertainties are large. Using our MCMC analysis we find $\sim$10 km s$^{-1}$ photon limited uncertainties, but the true uncertainty on each radial velocity measurement is likely larger due to systematic uncertainties related to our ability to continuum-normalize the spectra. When we change the way we continuum normalize the spectra (by dividing out polynomials of varying orders) we find the results remain consistent at approximately the 20 km s$^{-1}$ level. We adopt this value as the uncertainty on each radial velocity point in our analysis.

The function to which we fit the three RV points is of the form:
\begin{equation}
{\rm RV_i} = \gamma - K c_i \sin(\phi_i+\Delta \phi)
\label{eqn:RV1}
\end{equation}
where $i = 1,2,3$, $\gamma$ is the systemic velocity of the binary, K the orbital speed of the white dwarf, $\phi_i$ the nominal orbital phase computed from the photometry at the current epoch (see Sect.~\ref{sec:photometry}), $\Delta \phi$ a possible phase offset, and the $c_i$'s are the correction factors which take into account the finite exposure time, $\tau$, of the SDSS spectra:
\begin{equation}
c_i = \frac{P_{\rm orb}}{\pi \tau} \sin\left(\frac{\pi \tau}{P_{\rm orb}}\right) ~.
\label{eqn:correction}
\end{equation}
The $c_i$'s are independent of the orbital phase, except for a possible eclipse which we ignore.  Equation (\ref{eqn:RV1}) can be written in a form that is completely linear in three unknown parameters: $\gamma$, K\,$\sin \Delta \phi$, and K\,$\cos \Delta \phi$.
\begin{equation}
{\rm RV_i} = \gamma - (c_i \cos \phi_i) K \sin \Delta \phi - (c_i \sin \phi_i) K \cos \Delta \phi
\label{eqn:RV2}
\end{equation}
The RV values and correction factors that we use are given in Table \ref{tbl:RVs}.  We then solved for $\gamma$, K $\sin \Delta \phi$, and K $\cos \Delta \phi$, from which we find K and $\Delta \phi$.  

To the extent that we can fully accept the RVs determined from the SDSS spectra, we find $K \simeq 88 \pm 19$ km s$^{-1}$ and $\gamma \simeq 26 \pm 13$ km s$^{-1}$.  These fitted values of $K$ and $\gamma$ are also listed in Table \ref{tbl:RVs} along with the time of orbital phase zero and the corresponding offset from the phase projected back from the K2 observations.
 
\begin{table}
\centering
\caption{Radial Velocities from SDSS Spectra$^a$}
\begin{tabular}{lcccc}
\hline
\hline
Time & Expos. & RV$^b$ & $\phi$$^c$ & $c$$^d$  \\
(BJD$_{\rm tdb}$) & (seconds) & (km s$^{-1}$) & cycles & cycles \\
\hline
2452367.8107 & 900 & $+78.4$ & 0.750 &0.928 \\
2452367.8230 & 900 & $+86.6$ & 0.998 &0.928 \\
2452367.8369 & 1200 & $-36.2$ & 0.279 & 0.874 \\
\hline
\hline
Time ($\phi =0$) & $K$ & $\gamma$ & $\Delta \phi$$^e$ & \\
(BJD$_{\rm tdb}$)  & (km s$^{-1}$) & (km s$^{-1}$) & (offset)  & \\
\hline
2452367.8298 & $88 \pm 19$ & $26 \pm 13$ & $0.135 \pm 0.036$ & \\
\hline 
\label{tbl:RVs}
\end{tabular}

{\bf Notes.} (a) Spectral data taken from the SDSS spectral archive (Ahn et al.~2012). (b) RV values are from our cross-correlation analysis of the three spectra with a template white dwarf (see Sect.~\ref{sec:SDSS}) The uncertainty in each RV point is estimated to be 20 km s$^{-1}$ (see text). (c) The orbital phase computed from the ephemeris in Table \ref{tbl:binary}.  (d) The correction factors given by Eqn.~(\ref{eqn:correction}).  (e) Phase difference from the back-projected phases determined in 2016-2017.
\end{table}

\section{Ground-Based Photometry}
\label{sec:photometry}

\begin{figure*}
\begin{center}
\includegraphics[width=0.65\textwidth]{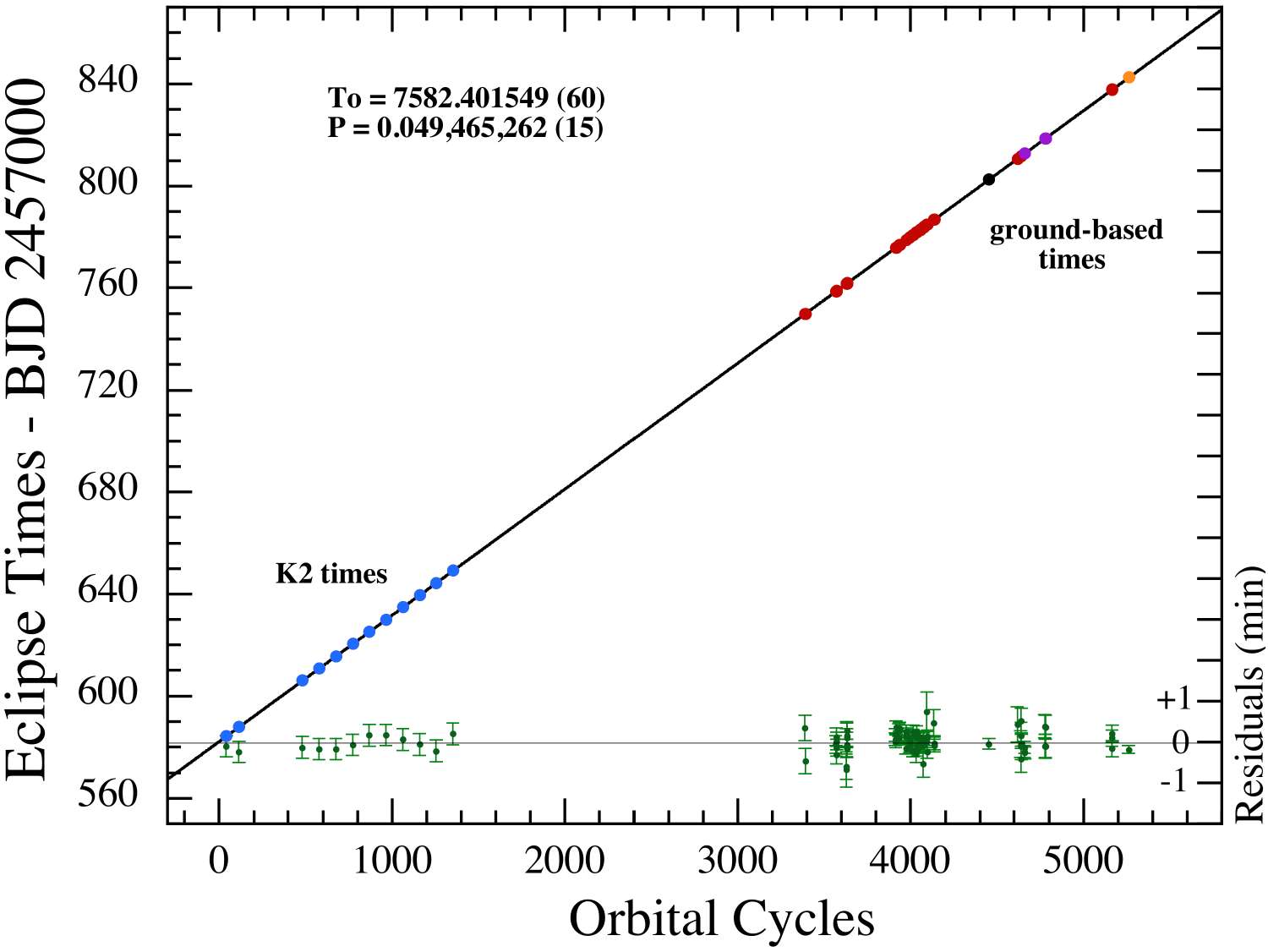} 
\caption{Summary of eclipse timing measurements of WD 1202-024.  The eclipse times are plotted vs.~the orbital cycle number.  The blue points to the lower left are from the K2 observations, while most of the points to the upper right (in red) were acquired with the Gary/Kaye facilities.  The black point is from the 1-m telescope of the South African Astronomical Observatory, the purple points are from the OMM 1.6-m data, and the orange point is from the Magellan observation. The residuals from a best-fit linear period are shown at the bottom with green points and error bars; the scale is on the right and is expressed in minutes.}
\label{fig:timing} 
\end{center}
\end{figure*}

We followed up the K2 observations of WD 1202-024 with ground-based photometric observations.  The ground-based data were acquired with amateur-operated 36-cm and 80-cm telescopes in Arizona, the 1-m telescope of the South African Astronomical Observatory, and the 1.6-m telescope at Observatoire Mont M\'egantic (`OMM') in Qu\'ebec.  We also obtained 15 minutes of photometry around the time of eclipse with the Magellan 6.5-m telescope\footnote{This paper includes data gathered with the 6.5 meter Magellan Telescopes located at Las Campanas Observatory, Chile.}.

The amateur-astronomer measurements in Arizona covered 61 individual orbital cycles during the 87-day interval from 2016 December 28 to 2017 March 24. The Gary observations conducted at the Hereford Arizona Observatory (HAO) used a 36-cm telescope, and the data were reduced by author BLG. The observatory, observing procedure and reduction process are described in Rappaport et al.~(2016).  The Kaye/Gary observations were conducted with the Junk Bond Observatory (`JBO') 80-cm telescope by author TGK of the Raemor Vista Observatory.  Image sets were calibrated and measured by BLG. The observatory, observing procedure and image reduction process are also described in Rappaport et al.~(2016).  The exposure times for the latter observations were either 30 or 40 seconds (with cadences of 42 and 54 sec).  During most nights of observation, approximately 4 complete orbital cycles were recorded. 

In addition to these observations we also acquired one observation covering nearly two orbital cycles on the night of 2017 Feb.~18 using the 1-m telescope at the South African Astronomical Observatory. These data were taken with a shorter integration time of only 8 s using the Sutherland High Speed Camera (`SHOC'; Coppejans et al., 2013) so as to better study the eclipse ingress and egress.  

Following on these observations, we obtained photometric observations covering 7 orbital cycles on the nights of 2017 Feb.~28 and March 6 at OMM.  These data were taken with the PESTO (Planetes Extra-Solaires en Transit et Occultations) camera on the 1.6-m telescope using 8-s cadence.   

All of the above observations utilized an `open' or `clear' filter which yields a bandpass largely dictated by the responses of the CCDs (roughly 0.4 to 0.9 $\mu$m).

Most recently we obtained 15 minutes of flux measurements using the LDSS3-C camera on the Magellan Clay telescope in Sloan i-band around the time of eclipse.  This was done specifically to further attempt to detect the companion star during total eclipse.  The observations were made on UT 30 March 2017 with exposures of 30 sec and a cadence of 60 sec. 

In order to determine a precise orbital period, we fit each eclipse for a mid-eclipse time.  These times are plotted vs.~the orbital cycle count in Fig.~\ref{fig:timing}.  In addition, we divided up the K2 photometric data into 12 discrete pieces, each of duration $\sim$5 days.  We determined the eclipse times for each of these 12 segments of K2 data.  These eclipse times are also plotted in Fig.~\ref{fig:timing}.  In all, the timing data span approximately 5220 orbital cycles or nearly 9 months.  All of our measured eclipse times, along with the observatories where the data were taken, are listed in Table \ref{tbl:eclipses}.  

The orbital period is 71.229,977(30) minutes, with a limit on orbital decay of $P_{\rm orb}/|\dot P_{\rm orb}| \gtrsim 2 \times 10^5$ years.

\begin{table*}
\centering
\caption{Eclipse Times for WD 1202-024}
\scriptsize{ 
\begin{tabular}{ccc c ccc}
\hline
\hline
Eclipse No. & BJD$_{\rm {\footnotesize TDB}}$-2457000 & Telescope$^a$ & ~~~ & Eclipse No. & BJD$_{\rm {\footnotesize TDB}}$-2457000 & Telescope$^a$\\
\hline
40  & 	584.3801 & K2  & & 4014  & 780.9550  & 80 \\
114  & 	588.0404 & K2  & & 4015  & 781.0047  & 80 \\ 
481  &	606.1942 & K2  & & 4032  & 781.8453  & 80 \\
578  &	610.9923 & K2  & & 4033  & 781.8951  & 80 \\
675  &	615.7905 & K2  & & 4034  & 781.9445  & 80 \\
772  &	620.5887 & K2  & & 4035  & 781.9937  & 80 \\
868  & 	625.3375 & K2  & & 4036  & 782.0433  & 80 \\
965  & 	630.1356 & K2  & & 4052  & 782.8347  & 80 \\ 
1062  &	634.9337 & K2  & & 4053  & 782.8843  & 80 \\
1158  &	639.6823 & K2  & & 4054  & 782.9337  & 80 \\
1255  &	644.4803 & K2  & & 4055  & 782.9832  & 80 \\
1352  &	649.2787 & K2  & & 4056  & 783.0327  & 80 \\
3388  &	749.9901  & 36  & & 4072  & 783.8237  & 80 \\
3389  &	750.0390  & 36  & & 4073  & 783.8736  & 80 \\
3567  &	758.8441  & 36  & & 4074  & 783.9230  & 80 \\
3568  &	758.8937  & 36  & & 4075  & 783.9725  & 80 \\
3569  &	758.9428  & 36  & & 4076  & 784.0220  & 80 \\
3570  &	758.9924  & 36  & & 4093  & 784.8634  & 80 \\
3571  &	759.0420  & 36  & & 4094  & 784.9123  & 80 \\
3628  &	761.8611  & 80  & & 4095  & 784.9616  & 80 \\
3629  &	761.9111  & 80  & & 4096  & 785.0114  & 80 \\
3630  &	761.9605  & 80  & & 4134  & 786.8913  & 80 \\
3631  &	762.0098  & 80  & & 4135  & 786.9404  & 80 \\
3911  &	775.8602  & 80  & & 4136  & 786.9899  & 80 \\
3912  &	775.9098  & 80  & & 4451  & 802.5714  & 100 \\
3913  &	775.9591  & 80  & & 4637  & 811.7719  &  80 \\
3915  &	776.0580  & 80  & & 4638  & 811.8211  &  80 \\
3932  &	776.8991  & 80  & & 4639  & 811.8710  &  80 \\
3933  &	776.9484  & 80  & & 4640  & 811.9207  &  80 \\
3934  &	776.9979  & 80  & & 4641  & 811.9698  &  80 \\
3935  &	777.0475  & 80  & & 4658  & 812.8106  & 160 \\
3972  &	778.8777  & 80  & & 4659  & 812.8601  & 160 \\
3973  &	778.9269  & 80  & & 4660  & 812.9095  & 160 \\
3974  &	778.9766  & 80  & & 4777  & 818.6974  & 160 \\
3975  &	779.0261  & 80  & & 4778  & 818.7465  & 160 \\
3992  &	779.8668  & 80  & & 4779  & 818.7960  & 160 \\
3993  &	779.9164  & 80  & & 4780  & 818.8458  & 160 \\
3994  &	779.9658  & 80  &  & 5164  & 837.8402  &  80 \\
3995  &	780.0152  & 80  & & 5165  & 837.8898  &  80 \\
4012  &	780.8560  & 80  & & 5166  & 837.9390  &  80 \\
4013  &	780.9057  & 80  & & 5261  & 842.6382  & 650 \\	
\hline
\label{tbl:eclipses}
\end{tabular}
}

\footnotesize{ {\bf Notes.} (a) Telescopes: K2 $\Rightarrow$ NASA K2 mission; ``36'' $\Rightarrow$ Hereford Arizona Observatory~36 cm; ``80'' $\Rightarrow$ Junk Bond Observatory~80 cm; ``100'' $\Rightarrow$ South African Astronomical Observatory (`SAAO') 100 cm; ``160'' $\Rightarrow$ Mont M\'egantic Observatory (`OMM') 160 cm; ``650'' $\Rightarrow$ Magellan 650 cm. The uncertainties in the eclipse times are approximately 10 s.}
\end{table*}

If we project the orbital phase backward in time by 14.3 years to the epoch when the SDSS spectra were taken, we can retain a coherent cycle count, but with a 1-$\sigma$ uncertainty in the phase of 0.040 cycles (corresponding to 2.8 minutes).  We note that the phase of the RV orbit misses the back projection of the photometric phasing by 9.6 minutes (see Table \ref{tbl:RVs}).  When taking in account both the uncertainty in the back projection as well as in the RV orbital phase determination (0.036 orbital cycles), the discrepancy is $9.6 \pm 3.8$ min., a 2.5-$\sigma$ discrepancy.  On the other hand, even the slightest component of $\dot P_{\rm orb}$ given above will allow the phases to match up.

\section{Modeling the Lightcurve} 
\label{sec:model}

\subsection{MCMC Physical Lightcurve Fitting}  
\label{sec:MCMC}

In this section we describe the fit to the lightcurves with a relatively simple physical model, while simultaneously imposing a set of supplementary constraints that allows for unique system solutions, modulo the uncertainties in the measurements and constraints.  We utilize an MCMC approach (see, e.g., Ford 2005; Madhusudhan \& Winn 2009, and references therein), and choose six parameters to fit via the MCMC procedure: the two constituent masses, $M_{\rm wd}$ and $M_{\rm com}$, the white dwarf radius, $R_{\rm wd}$, the orbital inclination angle, $i$, the age, $\tau$, of the binary (since its formation), and a thermal bloating factor, $f$, of the companion star. 

Armed with the MCMC parameters in each link, $T_{\rm eff}$ for the white dwarf, and an appropriate mass-radius relation for the low-mass companion star (discussed next), the shape of the primary eclipse can be computed.  We do that in 1-s steps around the orbit, and then convolve the results with the 30-40-s integration times of the JBO observations, or 8-s integration times for the SAAO and OMM observations.  In addition to the eclipse shape, we also need a constant term for the normalized-flux light curve, as well as two cosine terms to represent the illumination effect in the out-of-eclipse region.  For these, we follow Kopal (1959) and employ the following terms:
\begin{equation}
{\rm Illumination} \simeq -A \,\cos\left[\frac{2 \pi (t-t_0)}{P_{\rm orb}}\right] +B \,\cos\left[\frac{4 \pi (t-t_0)}{P_{\rm orb}}\right]
\label{eqn:illum}
\end{equation}
where $A$ and $B$ are parameters to be fit, and $t_0$ is the time of mid-eclipse, which we have already determined from our eclipse timing measurements.  The four linear parameters, $A$, $B$, the constant term, and a normalization factor for the eclipse profile are found via a simple linear regression for each step within the larger MCMC analysis.  

Because the lightcurve does not contain sufficient information to allow for the determination of the six MCMC adjusted parameters $M_{\rm wd}$, $M_{\rm com}$,  $R_{\rm wd}$, $i$, $\tau$, and $f$, we utilize six additional constraints or relations.

\noindent
{\em (i) RV amplitude from SDSS spectra}:
The two MCMC masses and the known orbital period determine the orbital separation, and the inclination angle allows for a prediction of the $K$-velocity of the white dwarf.  This predicted value is then ``tested'', via $\chi^2$ against the value inferred from the SDSS spectra, equivalent to imposing a Gaussian prior.  

\noindent
{\em (ii) Determination of $\log\,g$ from SDSS spectra}:
The MCMC mass and radius of the white dwarf are used to compute $\log \,g$.  This is compared to the value of $\log \,g$ inferred from the SDSS spectrum by Kleinman et al.~(2013), also equivalent to imposing a Gaussian prior on this quantity.  

\noindent
{\em (iii) Mass-Radius relation for an assumed H-rich companion}
Given the MCMC mass of the companion star, we compute its radius via a fitting formula that covers higher-mass brown dwarfs, `transition objects', and stars near the end of the main-sequence.  We derived a fitting formula for $R_{\rm com}(M,\tau)$ to represent stars of mass between 0.04 $M_\odot$ and 0.10 $M_\odot$ as a function of age, using our own models of low-mass stars and brown dwarfs.  All of the stellar models were computed using the Lagrangian-based Henyey method. The original code has been described in several papers (see, e.g., Nelson et al. 1985; Nelson et al. 2004) and has been extensively tested (Goliasch \& Nelson, 2015). We also reported results for low-mass main-sequence stars using this same code in Rappaport et al.~(2016).  The expression we have devised is:
\begin{eqnarray}
R_{\rm com}(M,\tau) & \simeq & f ( \alpha_1 + \alpha_2  M^{-1} + \alpha_3 \, \ell \tau + \alpha_4 M^{-2}   \nonumber \\
& & + \alpha_5 \, \ell \tau^2 + \alpha_6 \, \ell \tau \, M^{-1} +\alpha_7 M^{-3}   \nonumber \\
& & + \alpha_8 \, \ell \tau^3 + \alpha_9 \,\ell \tau^2 M^{-1} + \alpha_{10}\, \ell \tau  M^{-2} ) \nonumber \\
& & 
\label{eqn:RM}
\end{eqnarray} 
where $f \ge 1$ is an arbitrary multiplication factor that allows for the companion star to be thermally bloated by the external heating effects from the hot white dwarf, either currently or in the recent past when the WD might have been much hotter.  In this expression $M$ is the mass of the companion star, assumed to be a H-rich brown dwarf, `transition object', or lower main-sequence star over the range $0.04-0.10$ $M_\odot$, and $\ell \tau$ is the $\log_{10}$ of the stellar age in years since birth.  Eqn.~(\ref{eqn:RM}) holds for $8.5 \lesssim \log_{10}(\tau) \lesssim 10$.  The coefficients for this expression may be found in Table \ref{tbl:coeff}.

\begin{table}
\centering
\caption{Coefficients for Equation \ref{eqn:RM}}
\begin{tabular}{lc}
\hline
\hline
Coefficient & Value  \\
\hline
$\alpha_1$ & +7.90997319 \\
$\alpha_2$ & +0.02061960 \\
$\alpha_3$ & $-2.35535916$ \\
$\alpha_4$ & $-0.00078021$ \\
$\alpha_5$ & +0.23705829 \\
$\alpha_6$ & $-0.00279761$ \\
$\alpha_7$ & $-0.00001003$ \\
$\alpha_8$ & $-0.00775962$ \\
$\alpha_9$ & $-0.00020213$ \\
$\alpha_{10}$ & +0.00017943 \\
\hline
\label{tbl:coeff}
\end{tabular}
\end{table}

\noindent
{\em (iv) Mass-Radius relation for white dwarf at a given $T_{\rm eff}$:}
In Section \ref{sec:preferred} we will present a set of our own cooling curves (see Fig.~\ref{fig:cooling}) for recently unveiled low-mass white dwarfs that have lost their envelopes via Roche-lobe overflow. We have combined these models with those of Panei et al.~(2007), in order to increase the number of mass models, and show that for white dwarfs with the approximate $T_{\rm eff}$ of WD1202 (22,640 K; Kleinman et al.~2013), there is a relatively simple expression connecting the mass and the radius:
\begin{equation}
R_{\rm wd}(M_{\rm wd}) \simeq 0.00616 \,(M_{\rm wd}/M_\odot)^{-1.44} ~~R_\odot
\label{eqn:Rcool}
\end{equation}
(with an $\sim$8\% uncertainty in the theoretical values) which we use for masses up to $\simeq$ 0.50 $M_\odot$.  For higher masses, we simply set $R_{\rm wd} = 0.017\,R_\odot$.

\noindent
{\em (v) Radius of companion must be $<$ Roche-lobe radius}:
We impose the constraint that the chosen radius for the companion star must be less than, or equal to, its Roche-lobe radius, $R_L$.  

\noindent
{\em (vi) Matching the out-of-eclipse fit to the amplitude expressed analytically by Kopal (1959)}:  Here we utilize only the fitted amplitude $A$ in Eqn.~(\ref{eqn:illum}) (i.e., the illumination term at the orbital period) to match Kopal's (1959) analytic expression for $A$:
\begin{equation}
A \simeq -\frac{17}{48} \left(\frac{R_{\rm com}}{a}\right)^2 \left[1 + \frac{3}{4} \left(\frac{R_{\rm com}}{a}\right)\right] \frac{BC_{\rm com}}{BC_{\rm wd}}
\end{equation}
(see Eqn.~6 of Carter et al.~2010), where $a$ is the orbital separation, and the `BC' terms are the ratio of bolometric correction factors for the bandpass that produces the lightcurves.   

\begin{figure*}
\begin{center}
\includegraphics[width=0.70 \textwidth]{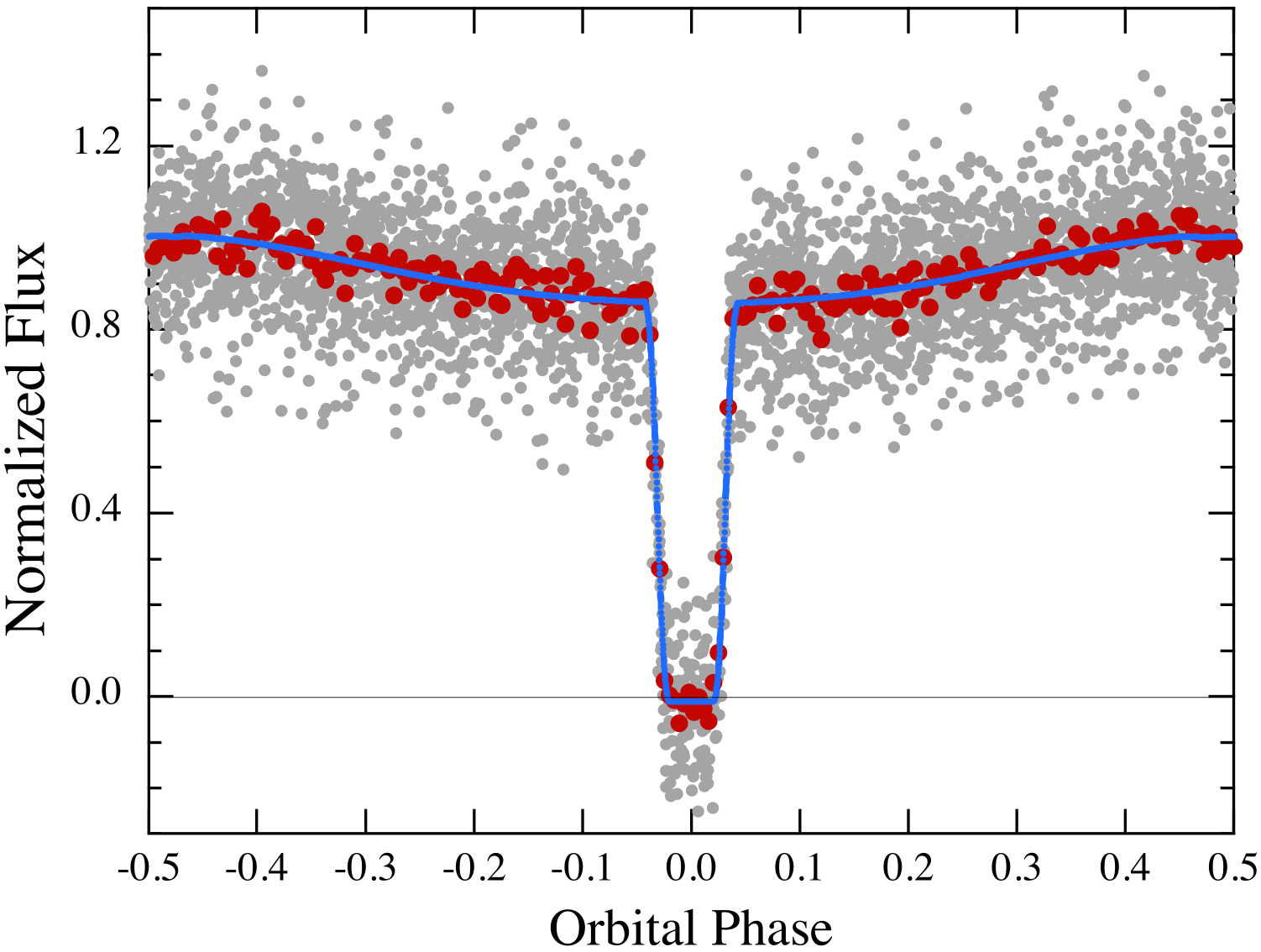} \vglue0.3cm
\includegraphics[width=0.70 \textwidth]{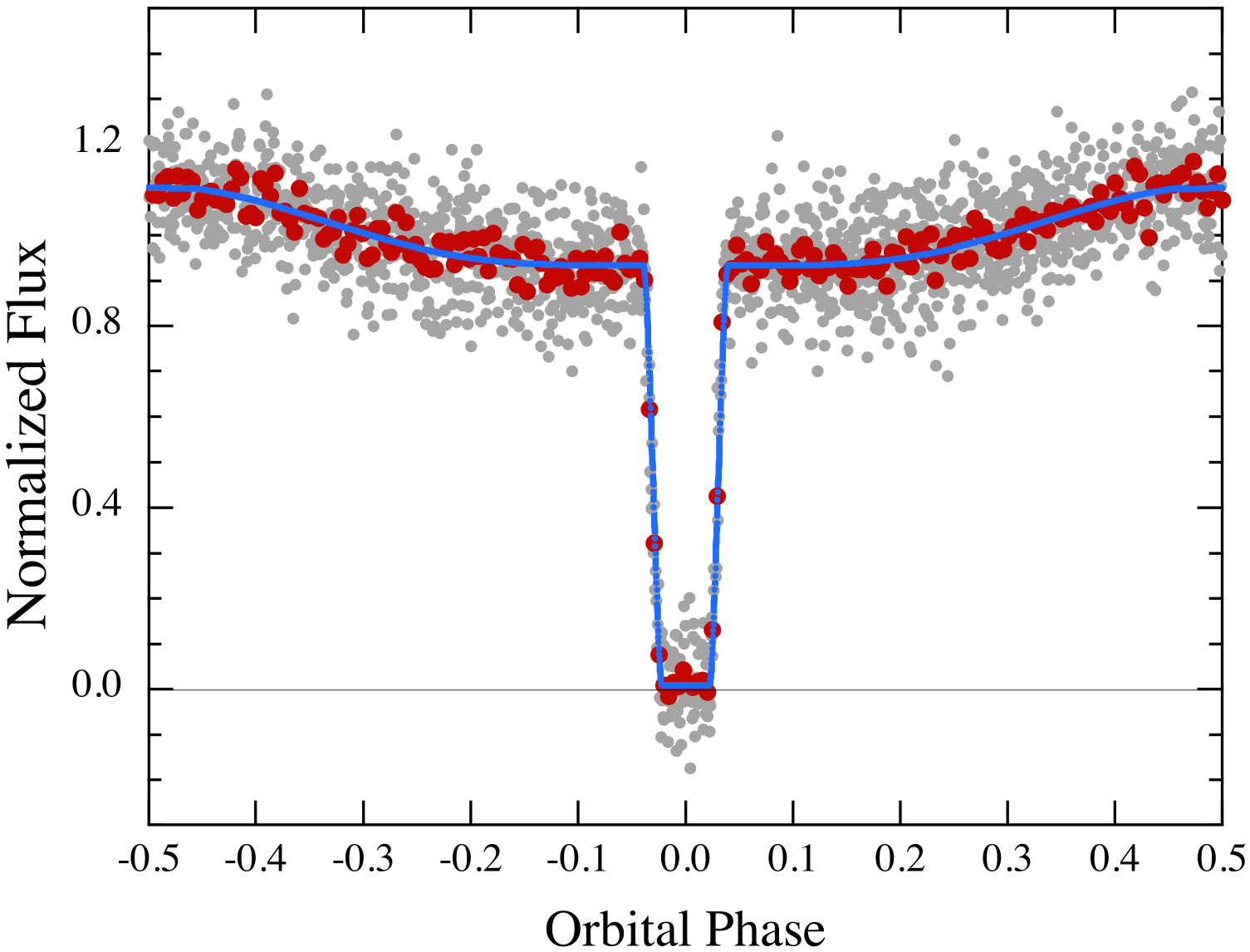} 
\caption{Fitted lightcurves for WD 1202-024.  Top panel is based on the Kaye/Gary data acquired with the JBO 80-cm telescope with the 30-s and 40-s exposure data sets combined (see Table \ref{tbl:eclipses}).  The bottom panel is from the OMM observations of 2017 Feb.~28 using 8-s cadence.  The gray and red points are the unbinned and binned data points, respectively, while the solid blue curve is the model fit discussed in Sect.~\ref{sec:MCMC}.}
\label{fig:fittedLC} 
\end{center}
\end{figure*}

For every given set of MCMC parameters, we compute $\chi^2$ from the fit to the light curve, and then add to this $\chi^2$ sum the contributions from constraints (i), (ii), (iv), and (vi) above. A decision about whether to accept the particular set of parameters or make a new selection is made based on the standard Metropolis-Hastings jump conditions (see, e.g., Ford 2005; Madhusudhan \& Winn 2009, and references therein).  After running the MCMC simulations, we find that constraints (ii) and (iv) are the most influential.

\begin{figure}
\begin{center}
\includegraphics[width=0.98 \columnwidth]{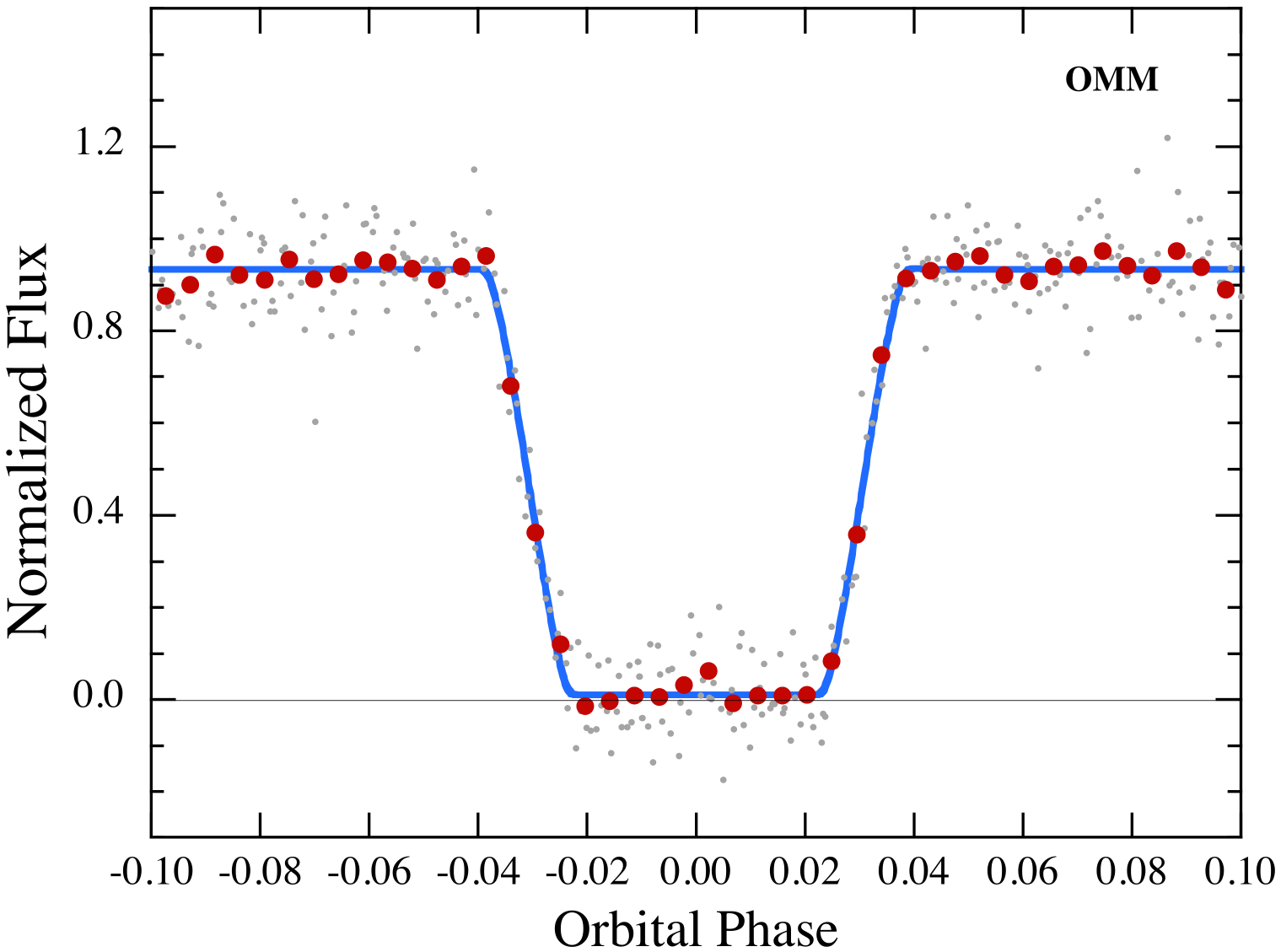}  \vglue0.3cm \hglue0.08cm
\includegraphics[width=0.98 \columnwidth]{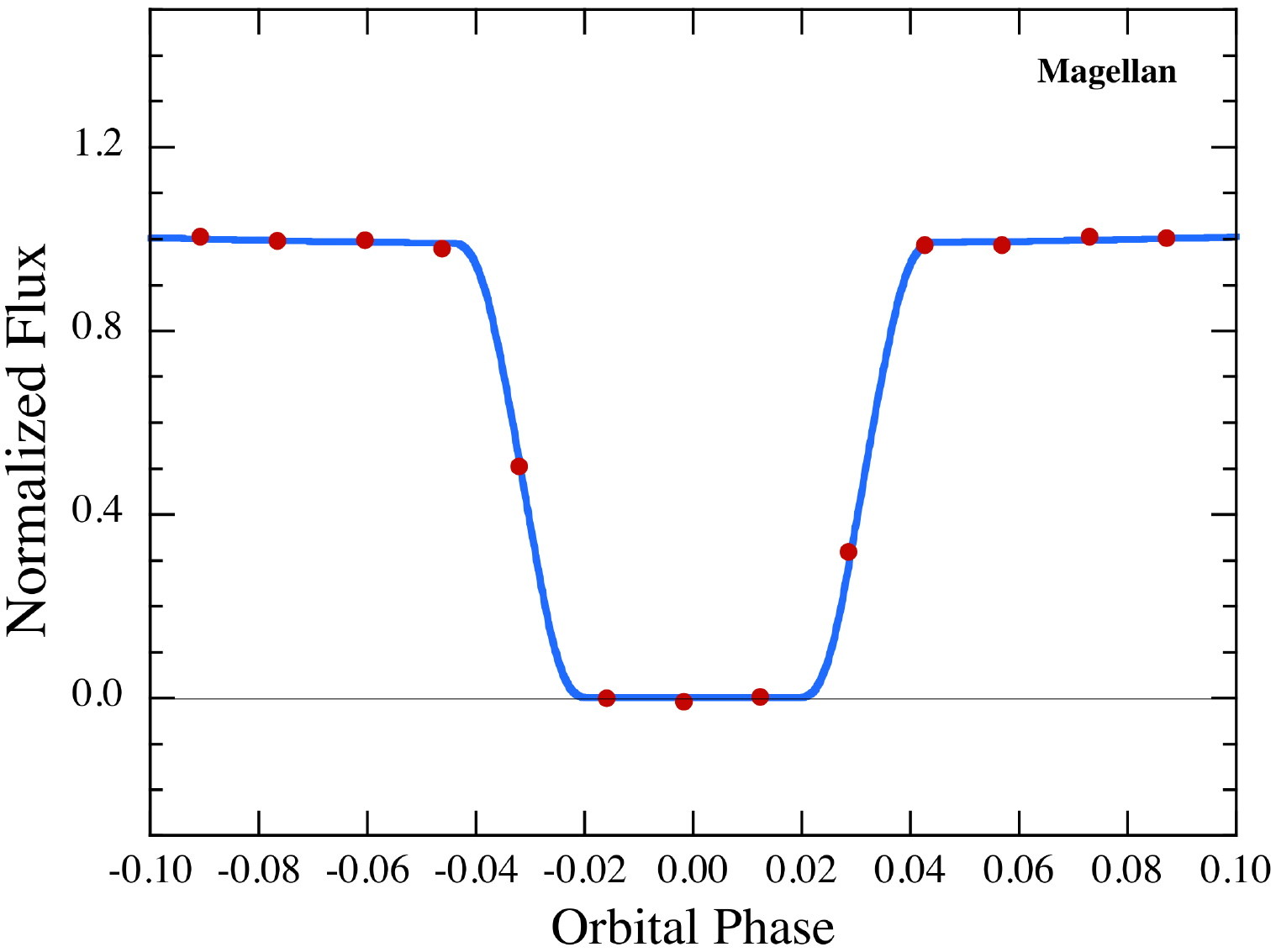}  
\caption{Fitted lightcurves zoomed in on the eclipse region.  Top panel: based on the data acquired with the OMM 1.6-m telescope of 28 Feb with 8-s cadence.  Bottom panel: from the Magellan Sloan-i band data of 30 March with 60-s cadence.  The solid blue curve is a model fit, using the same ingredients as the fits in Fig.~\ref{fig:fittedLC} (see Sect.~\ref{sec:MCMC}). The uncertainty on the Magellan data points are about half the size of the plotted points, and they are used to set a 3-$\sigma$ upper limit during eclipse of 1\% of the out of eclipse flux.}
\label{fig:fittedLC2} 
\end{center}
\end{figure}

Illustrative model fits to the lightcurves from the combined 30-s and 40-s JBO data sets and to the OMM 2017 Feb.~28 lightcurve are shown in Fig.~\ref{fig:fittedLC}. The grey points represent the individual flux measurements while the red points are the same data but binned and averaged in 200 points across the lightcurve.  Note the clear illumination effect of the companion star during the out-of-eclipse portion of the lightcurve.  There is no significant detection of flux during the total eclipse of the white dwarf, and only a marginal detection of the secondary eclipses (at the $\sim$1.5-$\sigma$ level).

In the top panel of Fig.~\ref{fig:fittedLC2} we show the region of the OMM lightcurve (from 28 Feb.) zoomed in near the eclipse.  The ingress and egress times are each about 1 minute in duration, which are well resolved with the 8-s cadence.  This portion of the lightcurve contains most of the information for determining the stellar radii and orbital inclination angle.  

In the bottom panel of Fig.~\ref{fig:fittedLC2} we show the same region of the lightcurve with the much sparser sampling but more precise photometry of the Magellan observations.  These data were acquired in the Sloan-i band.  The main thing we learn from these measurements is that there is no detectable flux during total eclipse of the WD down to a level of 1\% (3-$\sigma$) of the out-of-eclipse flux.  The equivalent magnitude is $\gtrsim 24$.

\subsection{Results of the Analysis}
\label{sec:results}

\begin{figure*}
\begin{center}
\includegraphics[width=0.45 \textwidth]{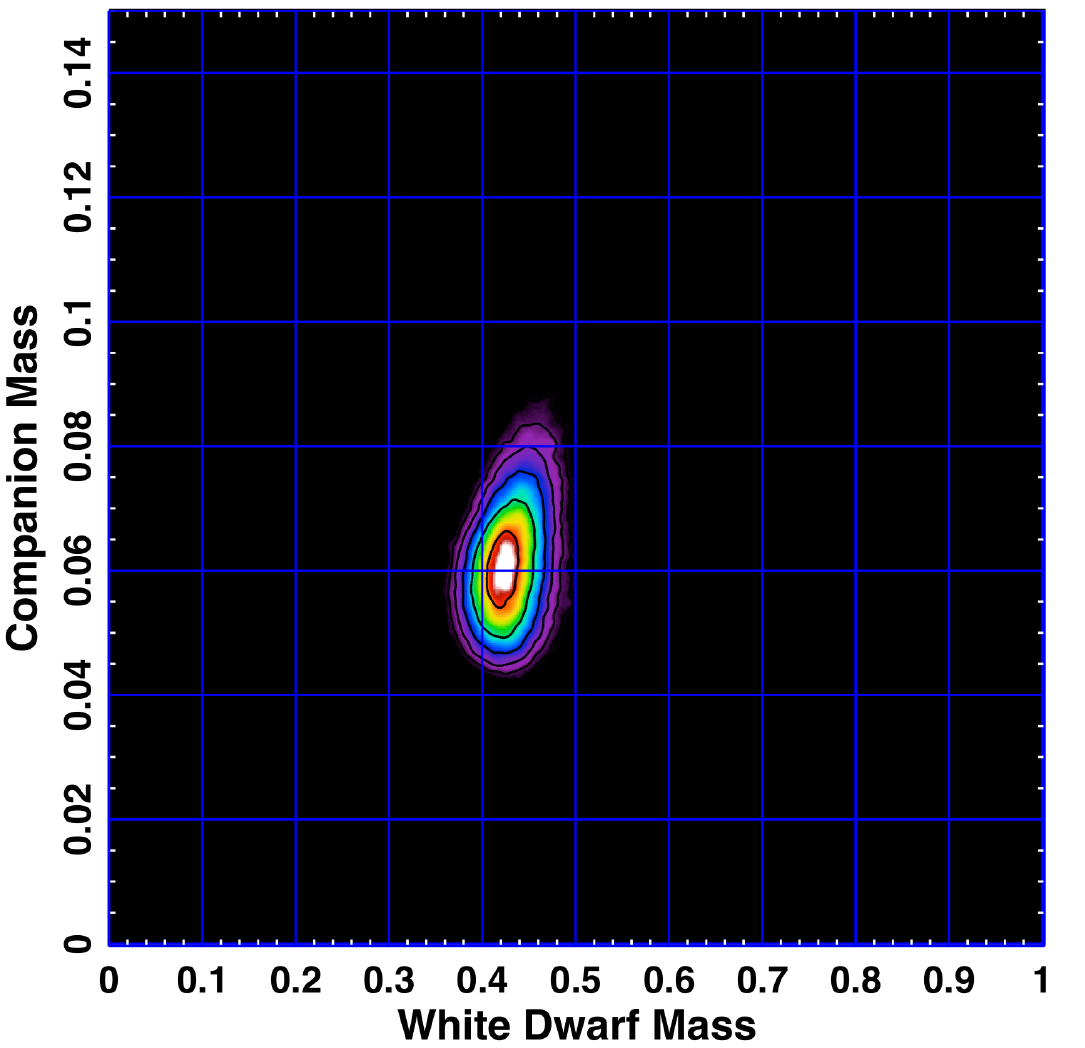} \hglue0.1cm
\includegraphics[width=0.447 \textwidth]{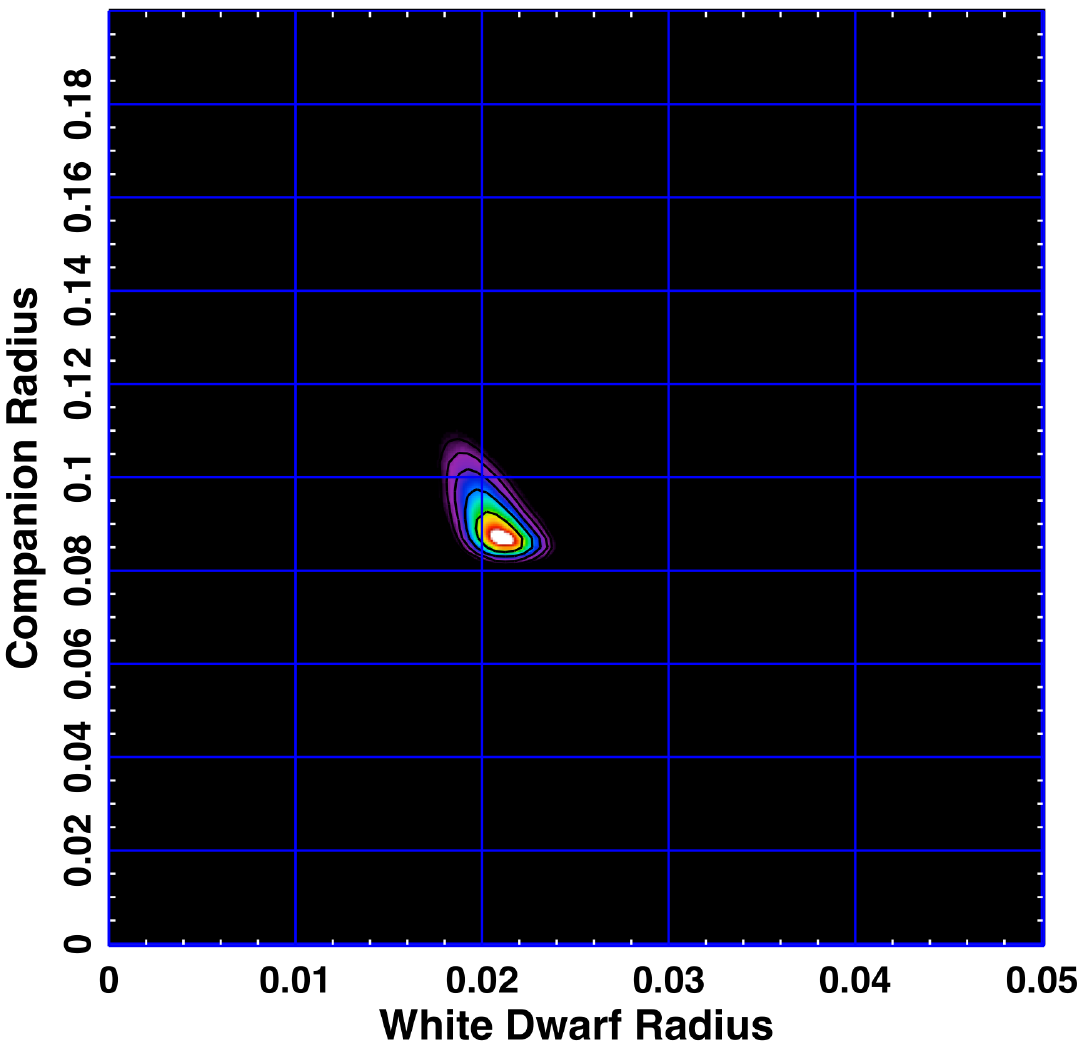} \vglue0.1cm \hglue0.15cm
\includegraphics[width=0.45 \textwidth]{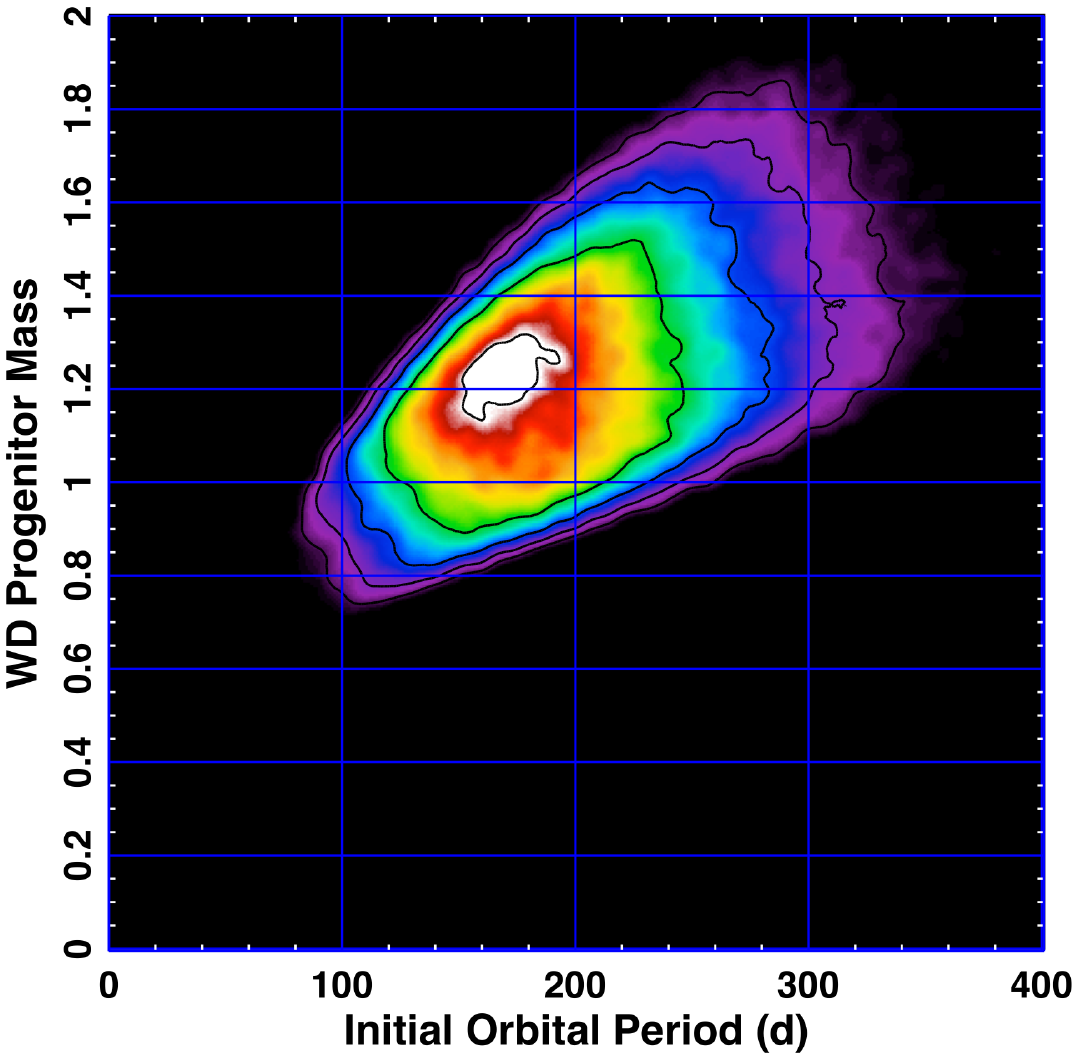} \hglue0.0cm
\includegraphics[width=0.45 \textwidth]{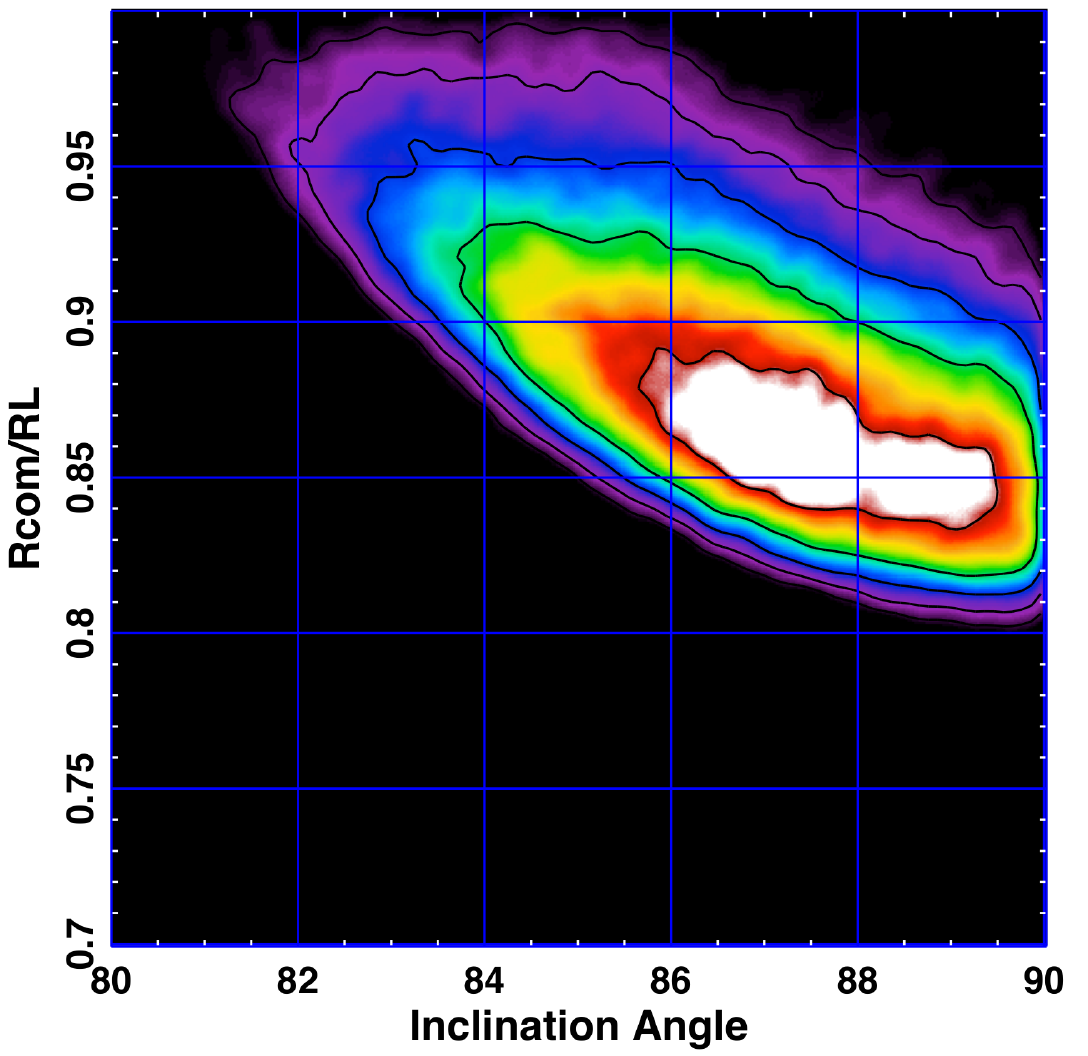} 
\caption{MCMC correlation plots for the WD 1202-024 system parameters based on the OMM Feb.~28 data set.  Clockwise from the upper left panel: $M_{\rm wd}-M_{\rm com}$, $R_{\rm wd}-R_{\rm com}$, $R_{\rm com}/RL - i$, and $M_{\rm p,init}-P_{\rm orb,init}$ planes.  Here $M_p$ is the WD progenitor mass and $P_{\rm orb,init}$ is the orbital period of the primordial binary at the onset of the common envelope.  Black contour curves are typically in steps of a factor of 2 in probability density.}
\label{fig:MCMCresults} 
\end{center}
\end{figure*}

The results of running a dozen chains with a million links each, are shown in Fig.~\ref{fig:MCMCresults}.  The parameter fits here are for the OMM Feb.~28 data set; however, fits to all the data sets are reported below in tabular form. The top left panel of Fig.~\ref{fig:MCMCresults} shows the correlation between $M_{\rm com}$ and $M_{\rm wd}$.  The white dwarf mass seems narrowly constrained to $0.415 \pm 0.028 \,M_\odot$, which makes it most likely a He WD.  The companion star has a mass of $0.061 \pm 0.010 \,M_\odot$ and a radius of $0.088 \pm 0.005 \,R_\odot$ (see upper right panel in Fig.~\ref{fig:MCMCresults}). This places the companion star near the boundary between brown dwarfs and the very tail end of the main sequence, i.e., a `transition object'.

From the upper right panel of Fig.~\ref{fig:MCMCresults} we see that the white dwarf has a radius that is narrowly constrained to be $0.021 \pm 0.001 \, R_\odot$.  Thus, it is substantially thermally bloated over its degenerate radius.  We utilize this in Sect.~\ref{sec:preferred}, along with the WD's $T_{\rm eff}$ of 22,640 K, to show that this is a relatively young WD and still quite thermally bloated.

In the bottom right panel of Fig.~\ref{fig:MCMCresults} we show the correlation between the orbital inclination angle and the Roche-lobe filling factor.  Inclination angles of $84^\circ - 90^\circ$ are strongly preferred.  Most interestingly the companion star appears to be somewhat, but significantly, underfilling its Roche lobe (with a filling factor of between 0.82 and 0.93; $< 0.97$ at 2-$\sigma$ confidence). As we discuss in Sect.~\ref{sec:origin} this points in the direction that this system is quite young, and that the mass transfer onto the WD has not yet commenced.  The bottom left panel in Fig.~\ref{fig:MCMCresults} will be discussed in Sect.\ref{sec:preferred}.

The properties of the WD 1202-024 binary are summarized in Table \ref{tbl:parameters}.  There we give the results of MCMC fits for each of the five individual data sets, as well as an average value across all the data sets.  We cite the average uncertainty in the parameter over the five data sets as a rough measure of the overall uncertainty. 

\begin{table*}
\centering
\caption{MCMC Derived Parameters of the WD 1202-024 Binary}
\begin{tabular}{lcccccc}
\hline
\hline
 & OMM-1 & OMM-2 & SAAO & JBO-1& JBO-2 & $\Leftarrow$ Observatory \\
 & 8-sec & 8-sec & 8-sec & 40-sec & 30-sec & $\Leftarrow$ Exposure \\
 & 1728 & 1728 & 1050 & 2000 & 1390 & $\Leftarrow$ Flux samples \\
 & 0.086 & 0.122 & 0.074 & 0.105 & 0.137 & $\Leftarrow$ RMS Flux \\
\hline
 Parameter  &  &  &  &  &   &  Average$^a$ $\Downarrow$ \\
 \hline
$a$ ($R_\odot$) & $0.447 \pm 0.009$  &  $0.443 \pm 0.009$  &  $0.449 \pm 0.008$  &  $0.438 \pm 0.009$  &  $0.444 \pm 0.010$  & $0.444 \pm 0.009$ \\
$M_{\rm wd}$ ($M_\odot$) & $0.424 \pm 0.030$  &  $0.411 \pm 0.029$ &  $0.431 \pm 0.027$  &  $0.393 \pm 0.027$  &  $0.414 \pm 0.028$  & $0.415 \pm 0.028$ \\ 
$R_{\rm wd}$ ($R_\odot$) & $0.021 \pm 0.001$  & $0.022 \pm 0.001$  &  $0.020 \pm 0.002$  & $0.023 \pm 0.002$  & $0.021 \pm 0.001$  & $0.021 \pm 0.001$ \\
$M_{\rm com}$ ($M_\odot$)  & $0.060 \pm 0.010$  & $0.062 \pm 0.010$ &  $0.064 \pm 0.008$  & $0.059 \pm 0.010$  & $0.062 \pm 0.010$ & $0.061 \pm 0.010 $ \\ 
$R_{\rm com}$ ($R_\odot$)  & $0.087 \pm 0.005$  & $0.088 \pm 0.006$  &  $0.088 \pm 0.004$  & $0.087 \pm 0.005$  & $0.089 \pm 0.005$ & $0.088 \pm 0.005$ \\ 
Thermal Bloating, $f$$^b$ & $1.055^{+0.060}_{-0.034}$ & $1.056^{+0.047}_{-0.033}$ & $1.063^{+0.071}_{-0.039}$ & $1.068^{+0.078}_{-0.045}$ & $1.074^{+0.063}_{-0.045}$ &  $1.063^{+0.064}_{-0.039}$  \\
$R_{\rm com}/R_L$$^c$ & $0.82-0.93$ &  $0.82-0.94$ & $0.82-0.89$ & $0.82-0.93$ & $0.83-0.94$ & $0.82 - 0.92$ \\
$i$ (deg) & $84 - 90$  & $83 - 90$  &  $87 - 90$  & $83 - 90$ & $84 - 90$ & $84-90$ \\
$\cos \omega_{\rm orb} t$ term & $-0.088 \pm 0.002$ & $-0.107 \pm 0.002$ & $-0.083 \pm 0.002$ & $-0.074 \pm 0.002$ & $-0.077 \pm 0.002$ &  $-0.086 \pm 0.013$$^d$ \\
$\cos 2\omega_{\rm orb} t$ term  & $+0.025 \pm 0.002$ & $+0.019 \pm 0.002$ & $+0.008 \pm 0.002$ & $+0.007 \pm 0.002$ & $+0.011 \pm 0.002$ & $ +0.014 \pm 0.008$$^d$ \\
\hline
\label{tbl:parameters}
\end{tabular}

{\bf Notes.} (a) Given that each data set has statistical uncertainties as well as systematic errors that may or may not be correlated with those of the other sets, for simplicity we  report straight averages of the five measurements and their mean uncertainties.  (b) The thermal bloating factor, $f$, is the value by which the companion star is found to be larger than its nominal radius given by Eqn.~(\ref{eqn:RM}).  These factors are already incorporated in the values given in the row for $R_{\rm com}$. (c) Roche-lobe filling factor.  The average 2-$\sigma$ upper limit is 0.97.  (d) Here we cite the rms of the scatter in the five values as the more appropriate uncertainty.
\end{table*}

\begin{table}
\centering
\caption{Summary Properties of the WD 1202-024 Binary}
\begin{tabular}{lc}
\hline
\hline
Parameter & Value \\
\hline
$P_{\rm orb}$$^a$ (min) & 71.229,977\,(30) \\
$P_{\rm orb}$$^a$ (day) & 0.049,465,262\,(20) \\
$\dot P_{\rm orb}/P_{\rm orb}$$^a$ & $5 \times 10^{-6} \,{\rm yr}^{-1}$ \\
$T_0$$^a$ (BJD) & 2457582.40155\,(6) \\
$K_{\rm wd}$$^b$ (km s$^{-1}$) & $88 \pm 19$ \\  
$K_{\rm wd}$$^c$ (km s$^{-1}$) & $59 \pm 7$ \\
$K_{\rm com}$$^c$ (km s$^{-1}$) & $396 \pm 10$ \\
$a$$^c$ ($R_\odot$) & $0.444 \pm 0.009$ \\
$M_{\rm wd}$$^c$ ($M_\odot$) & $0.415 \pm 0.028$ \\
$R_{\rm wd}$$^c$ ($R_\odot$) & $0.021 \pm 0.001$ \\
$T_{\rm wd}$$^d$ (K) & $22,650 \pm 540$ \\
$M_{\rm com}$$^c$ ($M_\odot$) & $0.061 \pm 0.010$ \\
$R_{\rm com}$$^c$ ($R_\odot$) & $0.088 \pm 0.005$ \\
$T_{\rm com, ss}$$^e$ (K) & $5250 \pm 375$ K \\
$T_{\rm com, heat}$$^e$ (K) & $4390 \pm 375$ K \\
$T_{\rm com, dark}$$^e$ (K) & $\lesssim 2450$ K \\
$i$$^c$(deg) & $84^\circ - 90^\circ$ \\
$L_{\rm bol,wd}$$^c$ ($L_\odot$) & $0.104 \pm 0.026$ \\
$L_{\rm v,com}/L_{\rm v,wd}$$^c$ & $\lesssim 1\%$ \\
${\rm age}_{\rm wd}$$^f$ (Myr) & $50 \pm 20$ \\  
$\tau$ (${\rm age}_{\rm binary}$)$^{c,g}$ (Gyr) & $3.4^{+2.7}_{-1.5}$ \\
$M_{\rm prim,init}$$^{c,g}$ ($M_\odot$) & $1.18 \pm 0.26$ \\
$P_{\rm bin, init}$$^{c,g}$ (days) & $155^{+60}_{-40}$ \\
\hline
\label{tbl:binary}
\end{tabular}

{\bf Notes.} (a) From the photometric observations (Sect.~\ref{sec:photometry}). (b) Based on our cross-correlation analysis of the data from three individual SDSS spectral exposures (see Sect.~\ref{sec:SDSS}).  (c) Based on the MCMC analysis of the system parameters (see Sect.~\ref{sec:MCMC}). (d) Taken from the SDSS spectra (Kleinman et al.~2013). (e) The subscripts ``ss'', ``heat'', and ``dark'', refer to the temperature at the substellar point of the white dwarf, the mean temperature on the heated hemisphere, and the effective temperature on the unheated side of the star (close to the companion's intrinsic temperature), respectively.  The cited uncertainties for the former two temperatures come from using an albedo of $0.25 \pm 0.25$.  (f) Inferred from the white dwarf cooling curves shown in Fig.~\ref{fig:cooling}. (g) The age of the binary since its primordial inception, the initial mass of the WD progenitor, $M_{\rm prim,init}$, and the initial orbital period of the primordial binary, $P_{\rm bin, init}$, are derived as part of the MCMC analysis utilizing a common-envelope prescription (see, Sect.~\ref{sec:preferred}).
\end{table}  

\section{The Origin of WD 1202-024}
\label{sec:origin}

\subsection{Evolutionary Scenarios}
\label{sec:evolve}

From the empirical evidence collected thus far for WD 1202-024, and its analysis, a picture emerges of a very short period binary containing a hot, thermally bloated white dwarf in orbit with a transition object relatively close to the end of the main sequence.  Based on our analysis, the companion star seems to be somewhat underfiillng its Roche lobe.  To our knowledge, there is no evidence that this system is now, or ever was, a mass-transferring CV (see, e.g., Ritter \& Kolb 2003).  In particular, there is no evidence in the eclipse profile or the spectrum indicating the presence of an accretion disk, and no flickering in the light curve (the latter would have also indicated the presence of an accretion disk).  Moreover, there is nothing in the historical record concerning a prior nova event at the source location (see, e.g., Shafter 2017). In this regard, WD 1202-024 is not detected in any of the WISE images\footnote{\url{http://irsa.ipac.caltech.edu/Missions/wise.html}} indicating no clear excess above the ordinary thermal continuum spectrum of the white dwarf.

In spite of the fact that this object does not appear to be a CV, its orbital period is as short as, or shorter than, the minimum orbital period for CVs containing H-rich donor stars.  However, shorter period CVs (e.g., AM CVns, with $P_{\rm orb}$ as short as $\sim$6 minutes; Ritter \& Kolb 2003) exist, but they are thought to harbor He-rich donor stars (e.g., Kalomeni et al.~2016; and references therein).  The majority of CVs are thought to have commenced their mass transfer phase with the donor star having a mass as high as perhaps 2 $M_\odot$ (e.g., Kalomeni et al.~2016).  As the orbital period slowly decreases, from perhaps 10 hours, the donor star loses mass while the orbit shrinks until the minimum period occurs near 72--75 minutes with a donor star mass of $\sim$0.05 $M_\odot$ (see, e.g., Rappaport et al.~1982; Rappaport et al.~1983; Howell et al.~2001).  The minimum period is reached when the core of the donor star is becoming degenerate, while at the same time its envelope is still somewhat thermally bloated.  

\begin{figure}
\includegraphics[width=0.48 \textwidth]{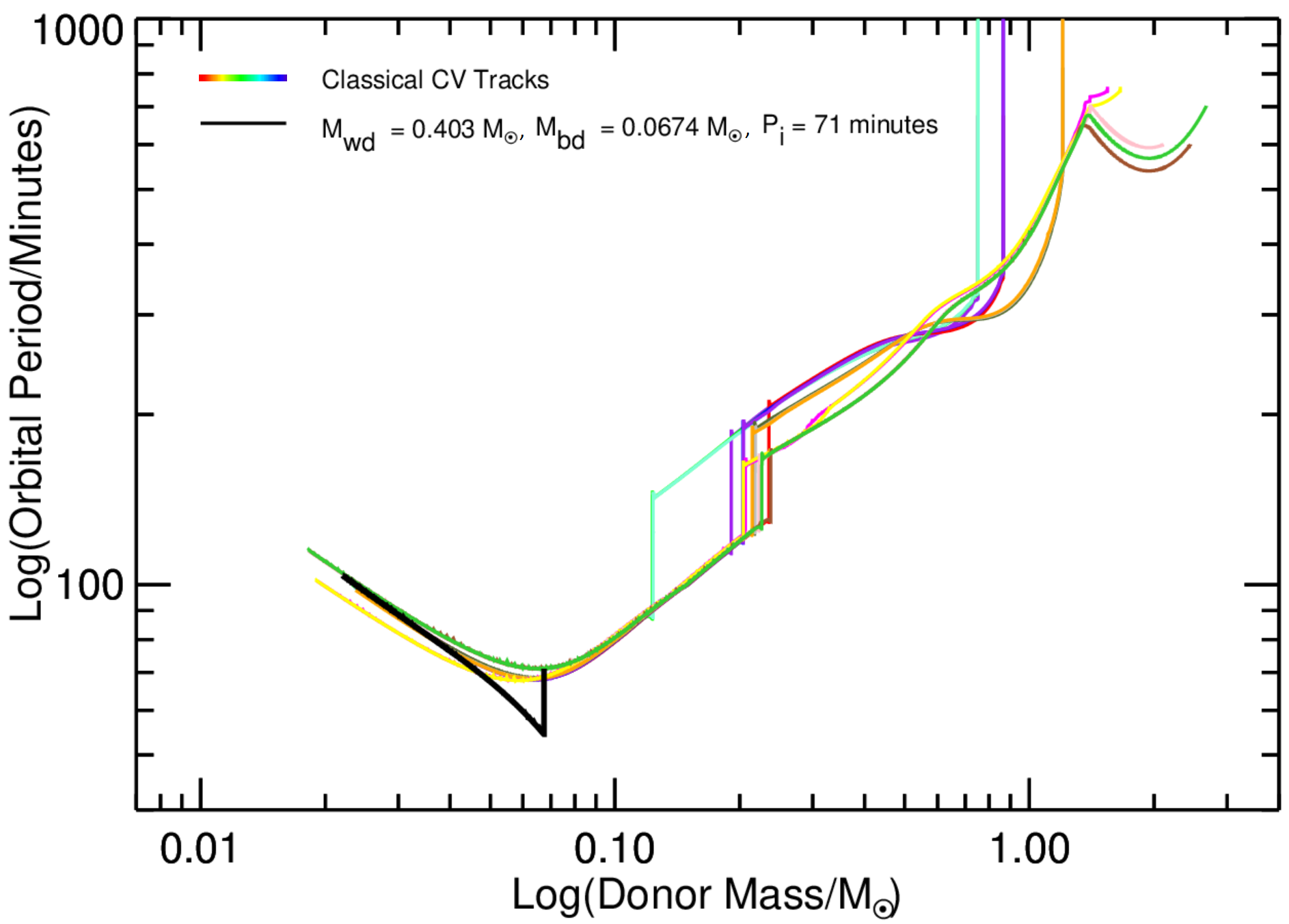}
\caption{Illustrative CV evolution tracks for 13 different sets of initial system parameters are shown as different colored curves (run with the MESA evolution code; Paxton et al.~2013, 2015; Kalomeni et al.~2016).  The initial white dwarf masses, donor-star masses, and orbital periods range from 0.6--1.4 $M_\odot$, 0.75--2.7 $M_\odot$, and 10--25 hrs, respectively.  Familiar features include the minimum orbital period at $\sim$70 min, as well as several `jumps' in period that represent the `period gap'.  The heavy black curve is the calculated evolution track of the current WD 1202-024 going forward in time from the present epoch.}
\label{fig:tracks} 
\end{figure}

To illustrate some of these standard CV evolution features, we selected 13 evolution tracks from the work of Kalomeni et al.~(2016), and plot them in Fig.~\ref{fig:tracks} in the $P_{\rm orb}-M_{\rm don}$ plane.  These systems all start with donor-star masses in the range of $\sim$0.75--2.7 $M_\odot$ and end up with those masses whittled down to 0.02 $M_\odot$.  They all have the property that they pass through a point in the $P_{\rm orb}-M_{\rm don}$ plane that is close to the current state of WD 1202-024.  Familiar features of these tracks include the minimum orbital period near 70 minutes and individual systems entering and leaving the period gap within the range of 2-3 hours. The ``U'' shaped feature near the start of several of the tracks is thermal-timescale mass transfer from the more massive donor stars to the white dwarf.  Systems that are transferring mass near a period of 70 minutes typically do so at a rate of $\sim$$3 \times 10^{-11} \, M_\odot$ yr$^{-1}$.

In this section we will adopt the working hypothesis that the WD 1202-024 system is {\em not} the result of the standard CV evolution depicted above, but rather has evolved directly to its current state from a relatively recent common-envelope phase (`CE').  During this CE phase the current companion star spiraled into, and ejected the envelope of the giant that was the progenitor of the current white dwarf.  This phase of its evolution occurred perhaps 50 Myr ago (see Sect.~\ref{sec:preferred}).  At the end of the CE phase, the orbital period was basically the same as its current 71 minutes, and the companion star was much as it is today, filling some 82\%--92\% of its Roche lobe. Specifically, the low-mass companion star would be nearly the same as it was in its primordial state.  In later subsections we return to reconsider the possibilities that (i) the companion star has evolved through a conventional CV phase and lost much of its mass, or (ii) the companion star filled its Roche lobe at the end of the common envelope phase and subsequently quickly lost a small fraction of its mass before settling into its current configuration.

\subsection{Preferred Scenario}
\label{sec:preferred}

In our preferred scenario, the system as we see it today formed directly as the result of a common-envelope phase, and the progenitor of the white dwarf has only relatively recently been stripped of its envelope, thereby unveiling the hot core of its progenitor.  To explore this scenario further, we start with the usual energy formulation for the common envelope evolution (see, e.g., Paczy\'nski 1976; Webbink 1984; Pfahl et al. 2003) which relates the initial binary orbital separation of the primordial binary, $a_i$, when the common envelope phase begins, to $a_f$, the final orbital separation once the common envelope has been ejected.

\begin{equation}
\frac{GM_p M_e}{\lambda r_L a_i}=\alpha \left[\frac{GM_c M_s}{2 a_f}-\frac{G M_p M_s}{2a_i}\right]
\label{eqn:CE1}
\end{equation}
where $M_p$ and $M_s$ are the masses of the primordial primary (the WD progenitor) and the primordial secondary star, respectively, and $M_c$ and $M_e$ are the masses of the core and envelope of the primary star (see, e.g., Taam et al. 1978; Webbink 1984; Taam \& Bodenheimer 1992; Pfahl et al.~2003). The parameter $\lambda^{-1}$ expresses the total energy of the primary star in units of $-GM_p M_e/R_p$, while $\alpha$ is an energy efficiency parameter for ejecting the common envelope.  Finally, the factor $r_L \equiv R_L/a_i$ is the dimensionless radius of the Roche lobe of the primary star when mass transfer commences.  

Typically, the second term in square brackets is negligible compared to the first term (a more detailed analysis can be found in Rappaport et al.~2015).  When this is dropped, we find:
\begin{equation}
\frac{a_f}{a_i} \simeq \frac{\lambda \alpha r_L}{2} \left(\frac{M_c M_s}{M_e M_p}\right)
\label{eqn:CE2}
\end{equation}
Through the use of Kepler's third law we can find the ratio of final to initial orbital periods:
\begin{equation}
\frac{P_f}{P_i} \simeq \left(\frac{\lambda \alpha r_L}{2}\right)^{3/2} \left(\frac{M_c M_s}{M_e M_p}\right)^{3/2} \left(\frac{M_p+M_s}{M_c+M_s}\right)^{1/2}
\label{eqn:CE3}
\end{equation}
We can then relate several of the masses in the above expression to the masses in the present binary (taken to be the `final state' of the common envelope process).  Specifically, $M_c \equiv M_{\rm wd}$, $M_s \equiv M_{\rm com}$, and $M_e \equiv M_p - M_{\rm wd}$.  

To uniquely solve the equations, we can make use of the core mass--radius relation for giants given by Eqn.~(5) of Rappaport et al.~(1995).  We then generalize their equation (7) to allow for (i) an arbitrary ratio of $M_p/M_s$, and (ii) a non-zero envelope mass for the giant:
\begin{equation}
P_i  \simeq  \frac{4 \times 10^4 \, m_{\rm wd}^{6.75}}{(1+4 m_{\rm wd}^4)^{3/2}} \frac{1}{r_L^{3/2}}\frac{1}{\sqrt{m_p+m_s}}~~{\rm days}
\label{eqn:PM}
\end{equation}
(see also Tauris \& van den Heuvel 2014) where we have used the above identity $M_{\rm wd} \equiv M_c$, and lower-case masses are expressed in solar units.  Here $r_L$ has the same meaning as in Eqns.~(\ref{eqn:CE2} and \ref{eqn:CE3}).

\begin{figure}
\includegraphics[width=0.49 \textwidth]{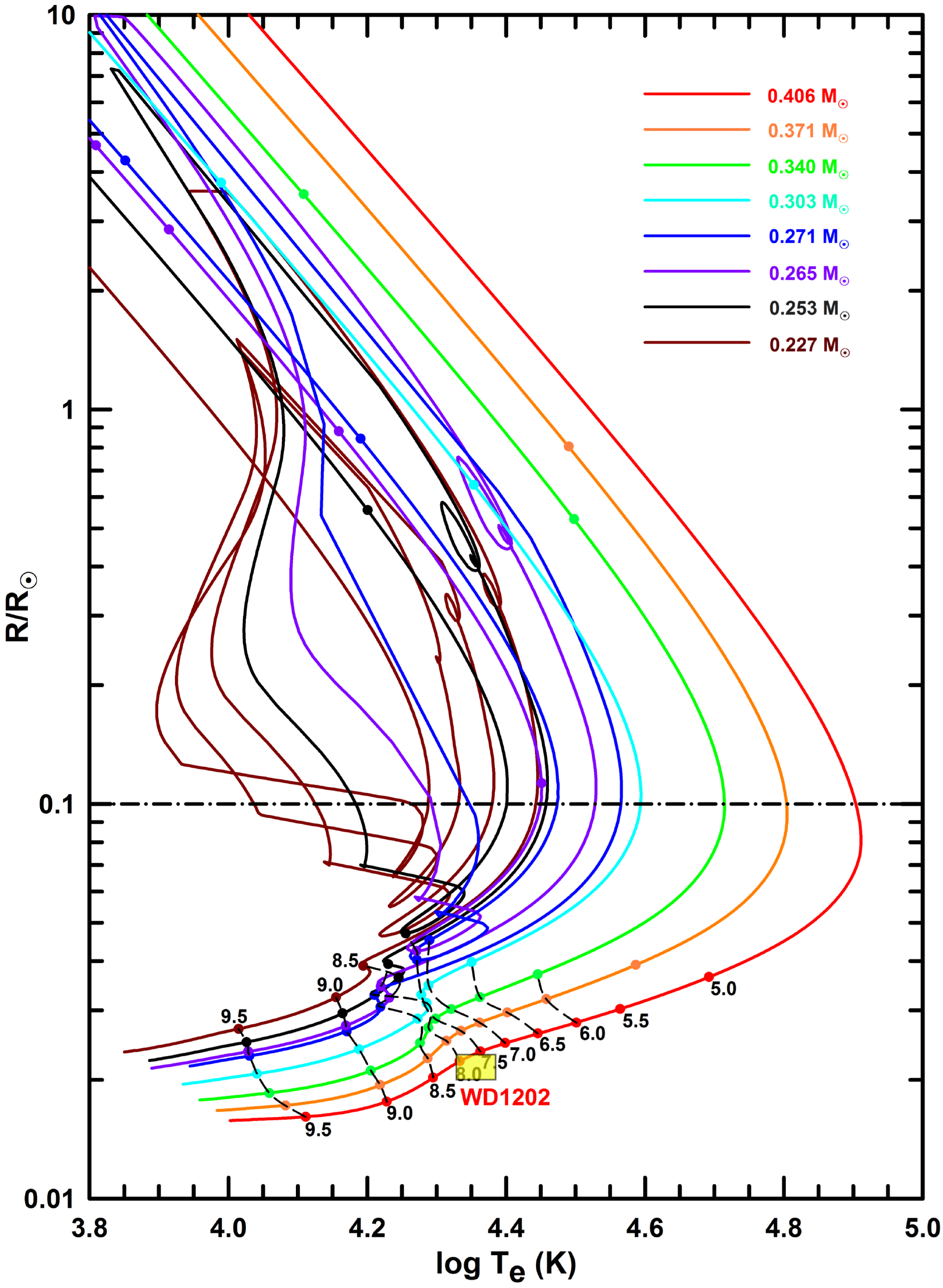}
\caption{Cooling curves for He white dwarfs whose envelopes were removed via {\em stable} Roche-lobe overflow. The curves are color coded by the final mass of the white dwarf. The Roche-lobe overflow should emulate the envelope removal of a common envelope albeit on a much slower time scale. We expect that immediately after the common envelope phase the proto-WD has a radius of $\lesssim 0.1 \, R_\odot$, and the cooling tracks start near the dot-dash horizontal line. Isochrones corresponding to the log of the cooling age (in yrs) extending from 6.0 to 9.5 are indicated by dashed lines and are labeled.  The boxed region (yellow rectangle) encloses the range of radii and $T_{\rm eff}$ that we are considering in this work.}
\label{fig:cooling} 
\end{figure}

If we now combine Eqns.~(\ref{eqn:CE3}) and (\ref{eqn:PM}), and identify $P_f$ with the current orbital period of 71.2 min., we can solve for the only unknown value, namely, $M_p$, i.e., the mass of the primordial primary star that led to the current system.  

The results of this type of calculation, directly computed during each MCMC link, are shown in the lower left panel of Fig.~\ref{fig:MCMCresults}.  There we can see the correlation between the mass of the white dwarf's progenitor star and the orbital period of the binary at the time mass transfer commenced, triggering a common-envelope phase.  The most likely initial orbital periods are $P_i \simeq 155^{+60}_{-40}$ days, and the corresponding masses of the primary star are $M_p \simeq 1.18 \pm 0.26 \, M_\odot$, though we note that there may be additional contributions to the error budget due to the uncertainties in the common-envelope parameter combination $\lambda \alpha$.\footnote{For the product of these two parameters we adopted randomly chosen values between 0.2 and 0.5 (see, e.g., Tauris \& Dewi 2001). For values of $\lambda \alpha $ as small as 0.1, the mass of the primary becomes too low for it to evolve during a Hubble time.}  This exercise mostly serves to demonstrate that it is quite plausible for a very low-mass star to successfully eject the envelope of a much more massive star without merging with its core (see also Politano 2004).  

Once the common-envelope phase has ended, the low-mass companion star is in a very tight 71-min orbit with the newly unveiled core of the primary, namely the He white dwarf that we see at the current epoch. Evolutionary cooling tracks for white dwarfs formed by having the primary star lose its envelope in binaries {\em via stable mass transfer}, are shown in Fig.~\ref{fig:cooling}.  We use the code devised by Nelson et al.~(2004) to compute the cooling evolution of the WDs. It is assumed that the red-giant progenitors are of solar metallicity and hydrogen abundance. After the proto-WDs start contracting, the compression of the hydrogen-rich envelopes for a narrow range of WD masses (0.21-0.29\,$M_\odot$) causes thermonuclear runaways (TNRs) to occur on their surfaces. This leads to a rapid expansion of the outer layers; more than one cycle can occur before the TNRs are quenched (see Fig.~\ref{fig:cooling}). Various cooling times are marked along the tracks in Fig.~\ref{fig:cooling} (the isochrones are denoted by dashed lines and are labeled in terms of the log of the cooling age in years).

Applying these results to WD1202-024, we find that for the first $10^5-10^6$ yr, or so, the evolving proto-WD would have a very large radius that would greatly overfill its Roche lobe in a 71-min binary. However, during a common envelope phase, which lasts much less time than $10^5$ yr (see, e.g., Sandquist et al.~1998), the H-rich tenuous envelope would be sufficiently extended that it would be quickly stripped away. By the end of the common envelope phase we would expect its thermally bloated radius to be no larger than $0.1 \, R_\odot$ due to the rapid stripping. A star of this latter radius could indeed be accommodated within the Roche lobe at a period of 71 min.  In any case, within a time of $\sim$$50 \pm 20$ Myr, the radius would have shrunk to a value close to that observed today, namely $R_{\rm wd}Ê\approx 0.021 \pm 0.001 \,R_\odot$ (the boxed region in Fig.~\ref{fig:cooling}). In fact, we used these later-time cooling tracks to derive the rough mass-radius relation that we employed in the MCMC evaluation of the system parameters, namely Eqn.~(\ref{eqn:Rcool}).

\subsection{Alternate Scenario}
\label{sec:alternate}

\subsubsection{Invoking a Prior Nova Explosion}

Even though it appears that WD 1202-024 is not now in Roche lobe contact and it is not transferring mass from the companion star to the WD, we should also consider the possibility that it had done so in the past.  Suppose that this is indeed a very old system, where the current companion star was initially much more massive, and it lost most of its envelope during a long ($\sim$Gyr) mass-transfer phase.  This raises several questions, including: (i) why is there now apparently no mass transfer occurring, and (ii) why is the white dwarf still very hot and thermally bloated?  Regarding the first of these, there is no evidence for any CV or dwarf nova activity reported from this object in the literature (see, e.g., Ritter \& Kolb 2003).  There are no evident emission lines in the spectrum due to an accretion disk (see Fig.~\ref{fig:SDSSspectrum}).  The photometric lightcurve of WD 1202-024 (see Figs.~\ref{fig:fittedLC} and \ref{fig:fittedLC2}) is symmetric about the eclipse as opposed to a number of eclipsing CVs undergoing mass transfer which exhibit distinctly asymmetric lightcurves due to the accretion disk hotspot (see, e.g., Fig.~1 of Littlefair et al.~2007 and Fig.~2 of Savoury et al.~2011).  Additionally, there is no prior X-ray activity reported for this source (e.g., from the Rosat All-Sky Survey\footnote{\url{https://heasarc.gsfc.nasa.gov/docs/rosat/rass.html}}), another possible signature of mass transfer in CVs.   

One possibility that might address both of the above issues is that the white dwarf underwent a large nova explosion some centuries, or even millennia, ago.  That could leave the WD in a hot and thermally bloated state, and perhaps slightly underfilling its Roche lobe (i.e., in a `hibernation' state; see, e.g., Shara et al.~1986; Kolb et al.~2001).  If an amount of mass, $\Delta m$ is ejected from the binary with the specific angular momentum of the white dwarf, then the fractional underfilling of the Roche lobe by the donor star after the nova explosion will be
\begin{equation}
\frac{R_L-R_{\rm com}}{R_{\rm com}}   \simeq \frac{4}{3} \frac{| \Delta m | }{M_{\rm wd}}  
\end{equation}
where we have taken $M_{\rm wd} \gg M_{\rm com}$.   Even for a rather large mass loss in a nova event of, say, $10^{-3} \, M_\odot$, this would result in a fractional underfilling of the Roche lobe by only 0.3\%.   

\begin{table*}
\centering
\caption{White Dwarfs in Close Detached Binaries with Brown-Dwarf or Transition-Object Companions}
\begin{tabular}{lcccccc}
\hline
\hline
 Object Name & $P_{\rm orb}$ & $M_{\rm wd}$ & $T_{\rm eff}$ & $M_{\rm com}$ & Eclipsing & Ref.  \\
  & (min) & ($M_\odot$) & (K) & ($M_\odot$) &  \\
\hline
GD 1400 & 600  &  0.67   &  11,600  &  0.06$^a$  & no &  1,2,3  \\
WD 0837+185 & 250  &  0.80  &  15,000  &  0.024$^a$  & no &  4  \\
SDSS J1557 & 136 & 0.45 & 21,800 & 0.063 & no & 5 \\
SDSS J1411 & 122 & 0.53 & 13,000 & 0.050 & yes & 6,7 \\
WD 0137-349 & 116  &  0.4   &  16,500  &  0.053   & no &  8,9,10  \\
NLTT 5306 & 102   &  0.44 & 7,756  & 0.053 & no & 11  \\
\hline
WD 1202-024 & 71  &  0.41  &  22,650   &  0.067  & yes &  this work  \\
\hline
\label{tbl:preCVs}
\end{tabular}

{\bf Notes.} (a) This value is $M_{\rm com} \sin i$. References: (1) Farihi \& Christopher (2004); (2) Dobbie et al.~(2005); (3) Burleigh et al.~(2011); (4) Casewell et al.~(2012); (5) Farihi et al.~(2017); (6) Littlefair et al.~(2014); (7) Beuermann et al.~(2013); (8) Casewell et al.~(2015); (9) Maxted et al.~(2006); (10) Burleigh et al.~(2006); (11) Steele et al.~(2013).
\end{table*}

The results of our analysis of the system parameters (see Table \ref{tbl:parameters}) indicate that the companion star is most probably filling 82\% to 92\% of its Roche lobe with little statistical probability in the MCMC distribution for a filling factor of $> 97\%$ (at the 2-$\sigma$ confidence level; see Table \ref{tbl:parameters} and Fig.~\ref{fig:MCMCresults}).  However, if our estimates are slightly off due to unknown systematic errors, then it is still conceivable that the companion is very close to filling its Roche lobe (i.e., at $\gtrsim 0.997$).  

\subsubsection{Implications of a Prior Nova Explosion}

WD 1202-024 is well above the Galactic plane (at $b^{II} \simeq 58^\circ$) with little visual extinction.  Thus, a nova at that location might well have been seen from the ground and reported, at least over the past century when better astronomical record-keeping began.  Yet there is no evidence for any historical nova at this location (see, e.g., Shafter 2017).  In spite of this lack of an historical record of a nova, it is possible that such an event occurred centuries or even millennia prior to the current epoch.  There have been numerous studies of the long-term cooling of the WD after nova explosions (see, e.g., Starrfield 1971; Prialnik 1987; Yaron et al.~2005; Bode 2004; Starrfield et al.~2004; and references therein).  It appears from these studies that the WD photospheres can cool to $\sim$50,000 K within about 30 years (i.e., within the span of time since the Palomar Sky Survey was first carried out).  However, cooling to values of $T_{\rm eff}$ of $\sim$23,000 K (comparable to WD 1202-024) could take anywhere from a few hundred to a few thousand years.  

We have checked these cooling times approximately using a Paczy\'nski (1983) one-zone shell model for nova explosions on a WD of 0.4 $M_\odot$, and find cooling times to 23,000 K for shell masses of 1, 3, and 10 $ \times 10^{-4} \, M_\odot$ of 500, 2300, and 5000 years, respectively.  In any of these cases, the cooling of the white dwarf over the past 60 years of photographic records of the object would have gone unnoticed.  

In the end, we tentatively conclude that WD 1202-024 was not a long-standing CV that happens to be turned `off' at the current epoch due to a past nova explosion or for any other reason.  Our main piece of evidence for rejecting this hypothesis must stand on our measurement that the companion star is currently underfilling its Roche lobe by at least 3-17\% and not merely $\lesssim 0.3\%$.

\section{Summary and Discussion}
\label{sec:concl}

We have found a short-period binary, WD 1202-024, consisting of a hot, thermally bloated white dwarf in orbit with a low-mass companion star that is of brown-dwarf size and mass.  Fortunately, there exist archival SDSS spectral data from which Kleinman et al.~(2013) deduced the $T_{\rm eff}$ and $\log \,g$ values for the white dwarf.  We were also able to use the sub-exposures  of the SDSS spectrum to extract three rough radial velocities over the course of an orbital cycle.  In addition, we have obtained ground-based follow-up photometric observations which have yielded lightcurves for this 19th magnitude object that are good enough to resolve the ingress and egress, as well as to readily detect the out-of-eclipse illumination curve of the heated companion star.  

When we analyze this lightcurve, coupled with the other constraints on the system, we are able to deduce the basic parameters of the binary.  The results are summarized in Table \ref{tbl:binary}.  The mass and radius of the white dwarf are $0.415 \pm 0.028 \, M_\odot$ and $0.021 \pm 0.001 \, R_\odot$, respectively, while those parameters for the low-mass companion star are $0.061 \pm 0.010 \, M_\odot$ and $0.088 \pm 0.005 \, R_\odot$.  The companion star fills between 82\% and 92\% of its Roche lobe, and $\lesssim 97\%$ with 2-$\sigma$ statistical confidence.

The white dwarf is extremely hot at 22,640 K, likely the result of having emerged from a common envelope some 50 Myr ago.  From the system geometry and the out-of-eclipse lightcurve, we infer that the substellar point on the companion star reaches a $T_{\rm eff}$ of 5400 K, while the {\em average} value of $T_{\rm eff}$ of the heated hemisphere is approximately 4390 K (see Table \ref{tbl:binary}).  We do not detect the companion star during total eclipse with a broad-band derived V-mag > 23, and in the Sloan i-band of 24th magnitude.  From that we infer that the temperature on the dark side of the companion is below $\sim$2500 K, or some $\gtrsim 1900$ K below that of the irradiated hemisphere.  Thus heat is not transported efficiently around to the dark side of the companion.  However, it would be very useful to follow up our study in the future with K-band observations, especially during the total eclipse.

Based on (i) the lack of evidence for mass transfer; (ii) the thermally bloated nature of the white dwarf; and (iii) evidence from our analysis that the companion star somewhat, but significantly, underfills its Roche lobe, we conclude that this system is not currently, nor has it ever been, a mass transferring CV.  Instead, it appears that the system has relatively recently emerged from a common envelope phase, and the current low-mass companion star is basically the original secondary star in the primordial binary.  If this is borne out by future studies of this system, then it provides a further demonstration that a star of brown-dwarf proportions can successfully eject the envelope of a much more substantive G star and end up in a very short period system, still underfilling its Roche lobe (see, e.g., Politano 2004; references in Table \ref{tbl:preCVs}).  There are six other known pre-CVs  with short periods and low-mass companions, and we summarize some basic information about these in Table \ref{tbl:preCVs} and references therein.  WD 1202-024 is then the shortest period pre-CV yet found (see, e.g., Zorotovic et al.~2011; Steele 2013; Casewell et al.~2015).  

There are several other short period binaries containing a white dwarf and an irradiated brown dwarf with system parameters comparable to those of WD 1202-024, but where mass transfer {\em is} clearly taking place.  These include SDSS J143317.78+101123.3 ($P_{\rm orb} \simeq 78$ min, $M_{\rm com} \simeq 0.057 \, M_\odot$; Hern\'andez Santisteban et al.~2016) and SDSS J150722.30+523039.8 ($P_{\rm orb} \simeq 66.6$ min, $M_{\rm com} \simeq 0.056 M_\odot$; Littlefair et al.~2007). In spite of the fact that they have similar system parameters, they are clearly in the process of transferring mass from the donor star to the white dwarf which leads to such phenomena as asymmetric lightcurves, emission from an accretion disk, and so forth.  This lends more confidence to the fact that, since these phenomena are not seen in WD 1202-024, this system is indeed detached.

If gravitational radiation is the only sink of angular momentum dissipation, we anticipate that WD 1202-024 will achieve Roche-lobe contact within $\lesssim 250$ Myr and that its orbital period will then be about 55 min.  It will then subsequently evolve to longer orbital periods (see black track in Fig.~\ref{fig:tracks}), in much the same way as do more typical CVs that have evolved starting with much more massive companion stars.

\vspace{0.3cm}
\noindent
{\bf Acknowledgements}

We are grateful to Paul Schechter for acquiring the Magellan Sloan-i band images. We thank Jules Halpern for facilitating the participation of J.\,R.\,T.~in this project, as well as for helpful discussions.  We thank the referee for extensive, insightful, and very helpful comments.  A.\,V.~is supported by the NSF Graduate Research Fellowship, Grant No.~DGE 1144152.  The amateur observers (B.\,L.\,G., T.\,G.\,K.) are self-funded so for them no institutional or tax-funded agencies need acknowledgement.  B.\,K.~acknowledge the support provided by the Turkish Scientific and Technical Research Council (T\"UB\.ITAK-112T766). J.\,R.\,T.~thanks the South African Astronomical Observatory for allocating telescope time, Patrick Woudt and David Buckley for help obtaining the time, the University of Cape Town Astronomy Department for their warm hospitality, and Mokhine Motsoaledi for help at the telescope.  This paper includes data collected by the K2 mission. Funding for the K2 mission is provided by the NASA Science Mission directorate. The Galaxy Evolution Explorer (GALEX) satellite is a NASA mission led by the California Institute of Technology. T.\,G.\,K.~thanks Beth Christie and Cheryl Healy for material support in JBO telescope operations and equipment.  L.\,N.~thanks NSERC (Canada) for financial support through the Discovery Grants program, F. Maisonneuve and E. Dubeau for their work on the stellar models, and Wen-Jian Chung and the staff at the Observatoire Astronomique du Mont-Megantic for their technical assistance.  Some computations were carried out on the supercomputers managed by Calcul Qu\'ebec and Compute Canada. The operation of these supercomputers is funded by the Canada Foundation for Innovation (CFI), NanoQuebec, RMGA, and the Fonds de recherche du Quebec -- Nature et technologies (FRQNT).











\bsp	
\label{lastpage}
\end{document}